\documentclass[fleqn,usenatbib]{mnras}

\usepackage{graphicx}	
\usepackage{amsmath}	
\usepackage{amssymb}	
\usepackage{comment}

\title[The formation of direct-collapse supermassive stars]{Radiation hydrodynamics simulations of the formation of direct-collapse supermassive stellar systems}

\author[S. Chon et al.]{
Sunmyon Chon,$^{1}$\thanks{E-mail: sunmyon.chon@utap.phys.s.u-tokyo.ac.jp}
Takashi Hosokawa,$^{2}$
and Naoki Yoshida,$^{1}$$^{3}$
\\
$^{1}$Department of Physics, School of Science, University of Tokyo, Bunkyo, Tokyo 113-0033, Japan\\
$^{2}$Department of Physics, Kyoto University, Kyoto 606-8502, Japan\\
$^{3}$Kavli Institute for the Physics and Mathematics of the Universe (WPI), Todai Institutes for Advanced Study
}

\date{Accepted XXX. Received YYY; in original form ZZZ}

\pubyear{2016}

\begin{document}
\label{firstpage}
\pagerange{\pageref{firstpage}--\pageref{lastpage}}
\maketitle

\begin{abstract}
  Formation of supermassive stars (SMSs) with mass $\gtrsim 10^4~M_\odot$ is a
  promising pathway to seed the formation of supermassive black holes in the early universe. 
  The so-called direct-collapse (DC) model postulates that such an SMS forms
  in a hot gas cloud irradiated by a nearby star-forming galaxy.
  We study the DC SMS formation in a fully cosmological context using three-dimensional
  radiation hydrodynamics simulations.
  We initialize our simulations using the outputs of the cosmological simulation
  of \cite{Chon+2016}, where two DC gas clouds are identified. 
  The long-term evolution over a hundred thousand years is followed from the formation of embryo
  protostars through their growth to SMSs. 
  We show that the strength of the tidal force by a nearby galaxy determines 
  the multiplicity of the formed stars and affects the protostellar growth.
  In one case, where a collapsing cloud is significantly stretched by strong tidal force,
  multiple star-disk systems are formed via filament fragmentation.
  Small-scale fragmentation occurs in each circumstellar disk,
  and more than 10 stars with masses of a few $\times 10^3~M_\odot$ are finally formed.
  Interestingly, about a half of them are found as massive binary stars.
  In the other case, the gas cloud collapses nearly spherically under a relatively
  weak tidal field, and a single star-disk system is formed. Only a few SMSs
  with masses $\sim 10^4~M_\odot$ are found already after evolution of a hundred thousand years,
  and the SMSs are expected to grow further by gas accretion
  and to leave massive blackholes at the end of their lives.
\end{abstract}

\section{Introduction}
Recent wide-field surveys discovered that a population of supermassive black holes (SMBHs) already
existed at redshift $z \gtrsim 6$ \citep{Willott+2010,Mortlock+2011,Wu+2015,Venemans+2016}.
Since the cosmic age at $z \simeq 6$ is about $0.8$~Gyr, the fact poses a challenge to the theory of SMBH formation.
One naive solution would be efficient growth of a seed BH with $\sim 10^2~M_\odot$ by gas accretion
over several hundred millions years, to attain barely the observationally
inferred mass $\gtrsim 10^9~M_\odot$ of the early SMBHs.
Obviously, the seed BH mass needs to be sufficiently large, but observationally confirmed
stellar-mass BHs via X-ray or gravitational wave (GW) emissions do not have masses as large as
100 $M_\odot$\citep[e.g.][]{Casares+2014,GW150914}.
Recent numerical simulations suggest that the mass distribution of Population III (Pop~III) stars 
can extend to $\sim 100~M_\odot$ \citep[e.g.][]{McKeeTan2008, Hosokawa+2011,Hosokawa+2016, Susa+2014,Hirano+2014,Hirano+2015}
and the most massive population leave equally massive BHs \citep[e.g.][]{Heger+2003}.
The remnant BHs may grow into SMBHs exceeding $10^9~M_\odot$ by $z \gtrsim 6$, if efficient accretion
with a near-Eddington rate, is maintained for about a billion years \citep[e.g.][]{Li+2007}.
However, it is already known that radiation feedback effects from both the stars and accreting BHs 
reduce the accretion rate far below the Eddington value
\citep[e.g.][]{Yoshida2006, Milos+2009, PR2011}.


Supermassive stars (SMSs) with $\gtrsim 10^5~M_\odot$ are thought to collapse via general relativistic instability
\citep[e.g.][]{Shibata+2002, Umeda+2016} to become massive BHs,
and hence may be promising ``seeds'' for the formation of early SMBHs \citep{BL2003, Hirano+2017b}.
Cosmological simulations suggest that the remnant BHs of early SMSs
can grow into SMBHs exceeding $10^9~M_\odot$ in several hundred millions years
via further mass accretion, and also by circumventing the stellar feedback effects 
\citep[e.g.,][]{DiMatteo+2012, Volonteri2010}. 


There are a few formation pathways of SMSs including the formation under supersonic gas streams \citep[e.g.][]{Hirano+2017b}
and a popular ultra-violet (UV) radiation driven collapse \citep{Chon+2016, Regan+2017}.
For the latter, consider that the Lyman-Werner (LW; 11.2 eV $\leq h \nu \leq 13.6~$ eV)
radiation emitted by nearby galaxies is strong
enough to destroy H$_2$ molecules in the cloud via photodissociation. If such a cloud is hosted by a massive dark halo
with $T_\text{vir} \gtrsim 8000~$K (the so-called atomic-cooling halo), the cloud collapses isothermally at $T \simeq 8000$~K
via atomic hydrogen cooling. Such a high temperature gas collapse induces rapid mass accretion onto a protostar
with a large rate of $\dot{M} \gtrsim 0.1~M_\odot~{\rm yr}^{-1}$ \citep[e.g.][]{Latif+2013}.
If the fast accretion continues, the stellar mass reaches $10^5~M_\odot$ within its lifetime of about a few million years.


The LW radiation intensity, usually normalized in units of $10^{-21} ~\mathrm{erg~s^{-1}Hz^{-1}cm^{-2}str^{-1}}$ ($J_{21}$),
is a key parameter to set the conditions for the so-called direct collapse (DC) of a primordial gas.
Various authors have investigated the critical $J_{21}$ for DC.
Recent studies suggest that $J_{21} \gtrsim 10^2$--$10^3$ is at least necessary \citep[e.g.][]{Omukai2001, Shang+2010, Latif+2014,
  Sugimura+2014}. 
The number density of the DCBHs has been also estimated by various 
authors \citep[e.g.][]{Agarwal+2012, Dijkstra+2014, Habouzit+2016}. 
They use galaxy formation models to investigate the distribution of $J_{21}$ at the locations of
atomic-cooling halos and count the
number density of the DC candidate halos in a volume of $\sim ~\mathrm{Mpc^{-3}} - \mathrm{Gpc^{-3}}$.
These previous studies, however, assume that all the halos that satisfy a few DC conditions
can actually bear SMSs. It remains unclear if 
the gas cloud in the DC halos can actually collapse and produce SMSs.


Earlier in \cite{Chon+2016}, we performed cosmological hydrodynamics simulations of the formation of
DC gas clouds.
We first used a semi-analytic model of early galaxy formation and traced
the star formation history of each halo to determine accurately 
the radiation intensity $J_{21}$ at the positions of neighboring atomic-cooling halos.
We then performed hydrodynamics simulations and found that a large fraction of
DC clouds do {\it not} collapse, because of disruption by the strong tidal force exerted
by the nearby galaxy.
Only 2 out of the 42 examined cases are successful in terms of DC.
The remaining important question is whether or not SMSs are formed in the `successful' cases.
We follow the subsequent protostellar accretion evolution in the present paper.


The final mass of an SMS is determined through dynamical interplay between the central
protostar(s), the circumstellar disk, and the infalling gas, where
gravitational fragmentation may occur.
Recent high resolution simulations show no significant fragmentation during the early collapse phase
of a DC cloud \citep{BL2003, Regan+2009a, Latif+2013, Inayoshi+2014, Choi+2015}.
In the later accretion phase after the birth of a protostar, however, the circumstellar
disk grows in mass and becomes gravitationally unstable to trigger
disk fragmentation \citep[e.g.,][]{Becerra+2015, Sakurai+2016}, possibly leading
to formation of a star cluster, rather than the formation of a single SMS. 


Radiative feedback from an accreting protostar is another key process,
which can limit the stellar mass growth by halting the accretion 
\citep[e.g.][]{McKeeTan2008,Hosokawa+2011}. 
Since the structure of an accreting protostar
under rapid accretion of $\gtrsim 0.1~M_\odot~\mathrm{yr}^{-1}$ 
is very different from that of a main-sequence star, with a very bloated envelope,
the UV feedback is thought to be weak for the DC case
\citep{Hosokawa+2012, Hosokawa+2013, Schleicher+2013}. 
Its effective temperature is only $T_{\rm eff} \simeq 5000$~K, with which 
the stellar UV emissivity remains small even when the stellar mass 
reaches $\sim 10^{4}$--$10^{5}~M_\odot$.
The UV feedback strength is intrinsically coupled to
the fragmentation described above
because individual protostars begin to contract if the accretion rate
falls below $\sim 10^{-2}~M_\odot~{\rm yr}^{-1}$. 


In this work, we study the protostellar evolution in the later accretion phase in the DC model.
We start our radiation hydrodynamic simulations from the final snapshots of
the two collapsing clouds found in \cite{Chon+2016}, and
follow the subsequent evolution for $\sim 0.1~$ Myrs.
We show that the tidal force exerted on the clouds critically affects the stellar mass growth. 
In one case, strong tidal field distorts the cloud and induces large-scale filament fragmentation.
The typical stellar mass is found to be as small as a few $\times~10^3~M_\odot$.
The other cloud experiences relatively weak tidal force, and stars with mass greater than $10^{4}~M_{\odot}$
are actually formed within 0.1 Myrs.
The latter case is a promising SMS formation process, whereas an interesting end-product
of the former case is massive star binaries. Both have important implications for direct
and indirect observations.
  
The rest of this paper is organized as follows. In Section \ref{sec_methodology}, we describe our numerical methods, setup, and initial condition. In Section \ref{sec_results}, we present the early evolution of the cloud collapse until the central density reaches $10^{13}~\mathrm{cm}^{-3}$, and then describe the evolution in the later protostellar accretion stage for the duration of $0.1$~Myr. 
In Section \ref{sec_radiation_feedback}, we investigate the role of the UV feedback in our simulations. The physical effects not included in our study are discussed in Section \ref{sec_discussion}. Finally, we summarize our results in Section \ref{sec_conclusion}.

\begin{figure*}
	\centering
	\includegraphics[width=15.5cm]{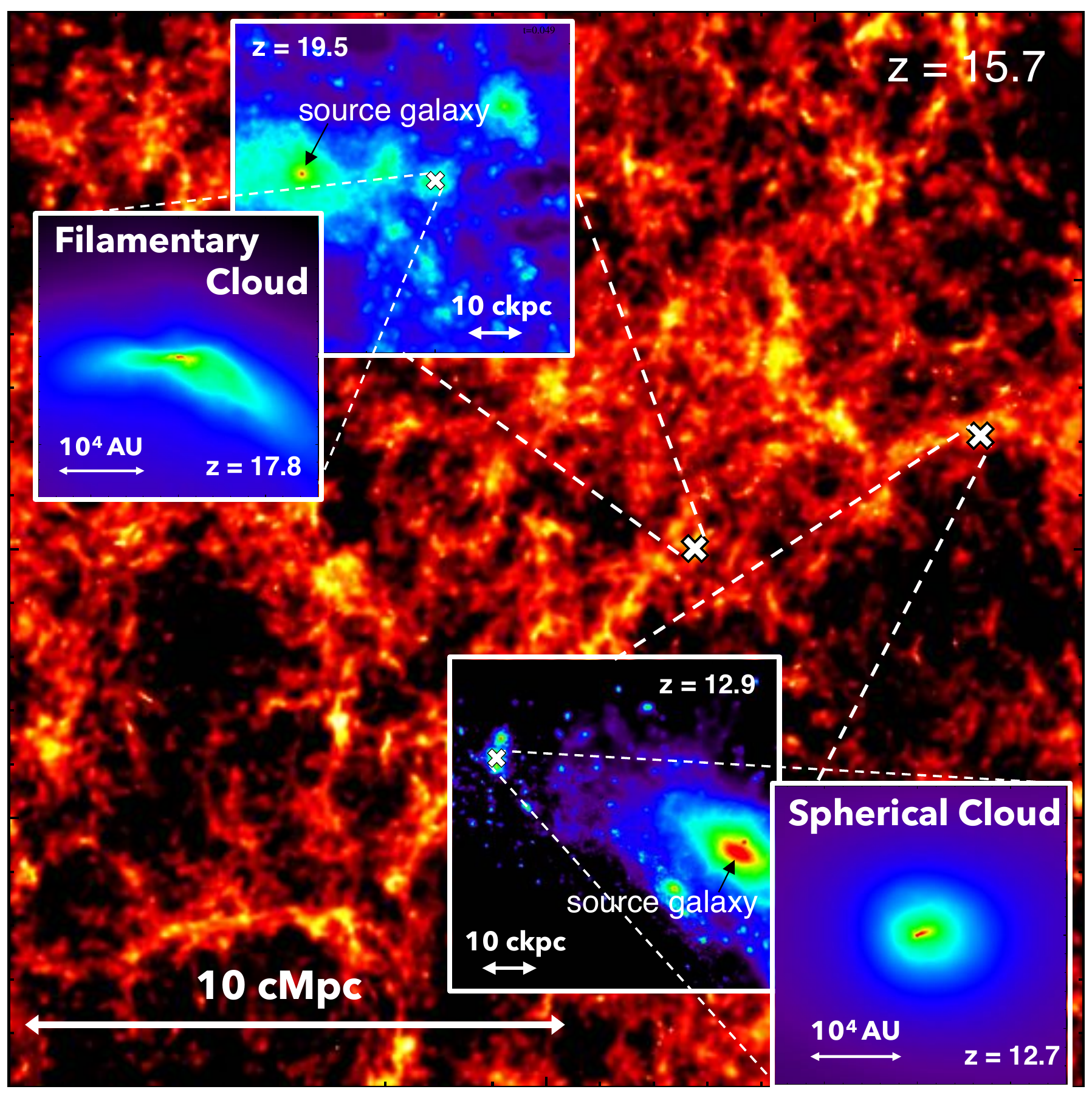}
	\caption{
Schematic view showing the locations of the spherical and the filamentary clouds. The clouds are taken from a large-scale cosmological simulation with the box size of $20~h^{-1}$Mpc on a side. In our previous study \citet{Chon+2016}, the cloud evolution has been followed until the central density reaches $10^8~\mathrm{cm}^{-3}$ during the early run-away collapse. We further follow the subsequent long-term evolution of the protostellar accretion for $\sim 0.1~$Myr in this paper.
	}
	\label{fig_LSS}
\end{figure*}

\section{Methodology} \label{sec_methodology}
We use the N-body + Smoothed Particle Hydrodynamics code
{\tt Gadget2} \citep{Springel2005} with the following extensions.
We implement sink particles to follow the protostar evolution. We use a stellar
evolution model to follow the radii (structure) and the luminosities of accreting
protostars. Finally, we implement a ray tracing method
to follow the radiation feedback from accreting protostars.
In this section, we describe our numerical method.

\subsection{Initial conditions}
\citet{Chon+2016} find that only 2 clouds collapse with the gas density reaching $10^8~\mathrm{cm}^{-3}$.
The clouds are assembled rapidly through mergers, and gravitational
collapse is triggered when their masses exceed a critical mass for DC.
Fig.~\ref{fig_LSS} summarizes the characteristics of these two clouds.
We label them as ``filamentary'' and ``spherical'' clouds. The filamentary cloud is
located close to the center of a massive galaxy with the separation of $\sim 100~$pc.
The nearby galaxy has a mass of $10^{10}~M_\odot$, which tidally distorts and elongates
the collapsing gas cloud (left top panel in Fig.~\ref{fig_LSS}).
The spherical cloud is located far from a massive galaxy with the
separation of $\sim 5~$kpc (right bottom panel in Fig.~\ref{fig_LSS}).
The LW radiation is mainly provided by less massive galaxy with
a mass of $\sim10^7~M_\odot$ at $\sim 400~$pc from the spherical cloud.
Thus the cloud can avoid the tidal deformation from the main galaxy
and is able to collapse almost spherically \citep{Visbal+2014}. 

To reduce the computational costs, we extract the region around the collapsing
cloud center when the central gas density has reached $10^8~\mathrm{cm}^{-3}$.
The extracted region has the radial extension of $\sim 10$ and $100~$pc from the cloud
center for gas and dark matter particles, respectively. We include the DM particles
in larger region than the gas particles, since the tidal field from
the nearby massive galaxy can be also important in the later accretion phase
of the collapsing clouds.

\subsection{Chemistry}
We include 9 primordial species ($\mathrm{e^-}$, $\mathrm{H}$, $\mathrm{H}^+$, $\mathrm{He}$, $\mathrm{He}^+$, $\mathrm{He}^{2+}$, $\mathrm{H}_2$, $\mathrm{H}_2^+$, and $\mathrm{H}^-$) and solve the chemical reaction
and the energy equation considering cooling process \citep{Yoshida+2003}.
Here, we omit the HD chemistry because the cloud temperature is so high
that the abundance of species including deuterium is negligible. 

We include an external radiation field which dissociates $\mathrm{H}_2$
and $\mathrm{H}^-$. The external LW intensity is set to be the value obtained by our semi-analytic
calculation in \cite{Chon+2016}. The intensities are $5000$ (filamentary) and $1000$ (spherical cloud)
in units of $J_{21}$ and almost constant throughout the calculation.
We assume that the radiation spectrum is described by a $10^4~$K black body as in \cite{Chon+2016}.
We do not consider attenuation of LW radiation
by the gas self-shielding since H$_{2}$ is mainly dissociated by the collision
with H atom (Section~\ref{sec:FUV_rad}).
We do not consider the ionizing radiation from the nearby massive
galaxy since it is attenuated by the intergalactic matter and
by the cloud envelope \citep{Chon+2017}.

Once protostars are formed, we also consider the local radiation feedback originating
from them. We implement only the local ionizing radiation which ionizes hydrogen atoms
by using a ray-tracing scheme (Section~\ref{sec_SEM}).
The local LW radiation is not included in our calculations for the following two reasons;
the intensity of the local LW radiation is smaller than or comparable to the external one,
and the H$_{2}$ dissociation rate by LW radiation is smaller than that by the collisional
dissociation with H atom in a hot gas \citep{Inayoshi+2012}. 
Thus, we expect the local LW radiation from accreting protostars
does not affect the thermal evolution of the collapsing clouds that are already exposed to
a strong external LW radiation.

We adopt an optically-thin Ly$\alpha$ cooling rate throughout the calculation,
even at densities greater than $\sim 10^4~\mathrm{cm}^{-3}$ where the gas actually
becomes optically thick to Ly$\alpha$ photons. In practice, this prescription regulates
the cloud temperature at $\sim 8000~$K, which resembles the realistic temperature
evolution that considers other H-cooling processes
\citep[e.g. 3p-2s transition, $\mathrm{H}^-$ continuum, see][]{Becerra+2015}.

Since we are interested in the stellar evolution for a long period of $\sim 0.1$--$1$ Myr,
we use a `hard' adiabatic equation state with $\gamma=5/3$ when the gas density exceeds
$n_\text{adib} =10^{13}~\mathrm{cm}^{-3}$. Otherwise, the increased gas density makes
the time step too short and the system's long-term evolution cannot be followed.
To see how this adiabatic prescription affects our results, we run test calculations
with changing $n_\text{adib}$ (see Section~\ref{sec:resolution} for details). 
In short, binaries with the smaller separations appear using the smaller threshold density. In our cases, however, the stellar radius becomes comparable to such a small separation with very rapid accretion (see Section~\ref{sec_SEM}). The binaries with separations of $< 100$~AU are merging away, and they do not survive in our simulations anyway.

\subsection{Particle splitting} \label{sec_splitting}
It is necessary to resolve the local Jeans length \citep[e.g.][]{Truelove1997,Nelson2006}.
We split a gas particle into 13 daughter particles following \cite{Kitsionas+2002},
when its density reaches $n=10^8$ and $10^{10}~\mathrm{cm}^{-3}$.
The initial mass of the gas particle is $1.6~M_\odot$ and the gas particle mass
with $n>10^{10}~\mathrm{cm}^{-3}$ is $9.4\times 10^{-3} ~M_\odot$.
This prescription allows us to resolve the Jeans length by more than ten times
the smoothing length of a gas particle \citep[e.g.][]{Stacy+2016}.

\subsection{Sink particles}
We implement a sink particle technique to emulate accreting protostars in a gas cloud.
We generate a sink particle when the local gas density
exceeds $n_\text{sink} = 5\times10^{13}~\mathrm{cm}^{-3}$. A sink particle
is characterized by two important parameters, a mass and a sink radius ($R$).
The region within $R$ around the sink particle is called the ``interaction region'',
inside which gas particles are accreted onto the sink particle.

As sink particle candidates, we first mark the gas particles with densities greater than $n_\text{sink}$.
To determine which gas particle should be replaced by a
sink particle, we impose the following additional criteria \citep{Hubber+2013}. 
\begin{enumerate}
\item Overlap criterion:\\
	The interaction region of a newly formed sink particle should not overlap with that of a preexisting sink particle $j$,
	\begin{equation}
	r_{ij} >  X_\text{sink} h_i + R_j,
	\end{equation}
	where $h_i$ is the smoothing length of the candidate gas particle $i$,
	$X_\text{sink}$ is an order of unity parameter which is set to be $4$,
	$r_{ij}$ is the separation between the candidate gas particle $i$ and the preexisting sink particle $j$, 
	and $R_j$ is the sink radius of the preexisting sink particle $j$. 
	Here, $X_\text{sink} h_i$ is the sink radius if the gas particle $i$ is replaced by the sink particle. 
\item Minimum gravitational potential:\\
  The particle should reside at the local gravitational potential minimum.
  We calculate the gravitational potential around the candidate particles.
  Then we examine whether they reside at the 
  potential minimum among other gas particles within $h_i$
  around the candidate particle. If not, we exclude the gas particle from the candidates.  
\item Hill criterion:\\
  The candidate gas particle (with index $i$) should have a
  sufficiently high density to contract under the disrupting tidal
  force from preexisting sink particle(s),
	\begin{equation}
	\rho_i > \rho_\text{Hill} \equiv \frac{3X_\text{Hill} a_{ij}}{4\pi G r_{ij}},
	\end{equation}
	where $G$ is the gravitational constant, $\rho_i$ is the density of
        the candidate gas particle $i$, $a_{ij}$ is the gravitational acceleration caused by the sink particle $j$, and $X_\text{Hill}$ is an order of unity parameter which is set to be $4$.
\end{enumerate}
The radius of the newly formed sink particle $i$ is set to be
$X_\text{sink} h_i$, which is $\sim 20~$AU in our calculations.

We assume that a sink particle accretes all the gas particles in the interaction region.
Note that the prescription likely overestimates the accretion rate \citep{Bate+1995, Hubber+2013},
because no gas particles are allowed to exist around the sink particles.
The surrounding gas particles outside the interaction region are also accelerated toward the sink particle
by the negative pressure gradient. 
We allow merging of sink particles, when the separation of two sinks becomes smaller
than the sink radius or the stellar radius ($R_*$, see Section~\ref{sec:star_supergiant}).
This is necessary because the giant protostars have radii comparable to the sink radius
when they are accreting the matter at a rate of $\sim 0.1 - 1~\mathrm{M_\odot~ yr^{-1}}$,
the typical values in the DC model \citep[e.g.][]{Hosokawa+2013}.
We conserve the mass and momentum of the sink particles before and after merging.
The angular momentum is also conserved, and the original orbital angular momentum is distributed to the gas just outside the interaction region. However, we do not exactly follow such an angular momentum redistribution because the detailed hydrodynamics during the merger is not fully resolved near the sink. Since each protostar experiences the merger event less than once during the simulations, the uncertainty will not significantly change our results.

\subsection{Stellar evolution model} \label{sec_SEM}
We employ an analytic model to estimate the radius, luminosity,
and effective temperature of an accreting protostar.
Our model reproduces well the evolution of the protostar properties 
for various accretion histories. The model has been tested with more detailed
calculations that solves the stellar interior structure equations
\citep[e.g.][]{Hosokawa+2013,Sakurai+2015}. In this study, we adopt
an even simpler model with
the following two evolutionary stages: (1) the supergiant protostar phase,
and (2) zero-age main sequence (ZAMS) phase. 

\subsubsection{Supergiant protostar phase} \label{sec:star_supergiant}
\cite{Hosokawa+2012,Hosokawa+2013} show that, with a high accretion rate
exceeding $\dot{M}_{\rm crit} = 0.04~M_\odot~{\rm yr}^{-1}$,
the evolution of the stellar radius is well approximated by the following relation
\begin{equation} \label{eq_rstar}
R_* = 38~\mathrm{AU} \left ({\frac{M_*}{1000~M_\odot}} \right )^{1/2},
\end{equation}
which means that the star has a very large radius and remains bloated as the stellar mass increases, independent of different accretion rates. During this stage, the stellar effective temperature is almost locked at $\simeq 5000$~K owing to very strong temperature dependence of H$^-$ opacity \citep[e.g.][]{Hayashi1961}. The protostar only emits a small amount of ionizing photons with such a low effective temperature. The resulting UV feedback is thus too weak to disturb the accretion flow \citep[e.g.][]{Hosokawa+2016}. Even if the accretion rate falls below $\dot{M}_{\rm crit}$, the protostar remains bloated for about ten times the Kelvin-Helmholz (KH) timescale \citep{Sakurai+2015}, 
\begin{equation} \label{eq_tKH}
t_\text{KH, surf} = 1000~\mathrm{yr} \left ( \frac{M_*}{500~\mathrm{M}_\odot} \right )^{1/2}.
\end{equation}

As an input to our analytic model, we estimate the accretion rate onto the protostar by averaging
instantaneous rates onto the sink particle over every 30 years.
We assume that the protostar is in the supergiant phase described by eq. \eqref{eq_rstar}
when the estimated accretion rate is higher than the critical value $\dot{M}_{\rm crit}$.
When being in the supergiant phase, a protostar emits little
amount of ionizing photons.
The protostar remains in the supergiant phase unless the accretion rates
fall below $\dot{M}_{\rm crit}$ for a duration longer than $t_{\rm KH, surf}$.

\subsubsection{Zero-age main sequence phase}

When the above conditions for the supergiant phase are not satisfied, the protostar begins to contract
owing to the radiative energy loss, leaves
the supergiant phase \citep{Sakurai+2015}, and enters the so-called KH contraction phase.
The stellar effective temperature rapidly rises up to $\sim 10^5$~K during KH phase.
After the contraction over a KH timescale, the accreting star turns to follow
the mass-radius relation of the ZAMS stars.

For our analytic model, we simply assume that the protostar is on the ZAMS if not in the supergiant phase.
We adopt our previous results of the stellar evolution calculations for the ZMAS stars,
i.e., the radius, luminosity, and effective temperature are tabulated as functions of the stellar mass.
We also assume that the star quickly returns back to the supergiant phase once
the accretion rate exceeds $\dot{M}_{\rm crit}$.

Since we ignore the finite KH timescale for the supergiant star to contract toward the ZAMS star,
our above prescription overestimates the stellar emissivity of the ionizing radiation.
As we will see in Section \ref{sec_radiation_feedback}, however, the ionizing photons
hardly affect the accretion flow in our simulations even with our simplified stellar model.

\subsection{Transfer of ionizing radiation} \label{sec_rt}
We calculate photo-ionization of neutral hydrogen and the resulting heating
using a ray tracing scheme proposed by \cite{Susa2006}.
The optical depth of ionizing radiation, $\tau_\text{UV}$,
from a light source to the particle $i$
is evaluated as the sum of the local optical depth, $\sum_j \mathrm{d}\tau_{\mathrm{UV},j}$.
Here, $\mathrm{d}\tau_{\mathrm{UV},j}$ is the optical depth from the particle $j$
to a particle located at the upstream of particle $j$. 
We then calculate the photon number which is locally consumed by the interaction with the neutral hydrogen.
Since the optical depth of one SPH particle is often large,
we use the so-called photon conserving method \citep{Kessel-Deynet+2000,Abel+1999},
where the photo-ionization rate $k$ and the photo-heating rate $\Gamma$ are
given by \citep{Rybicki&Lightman1979}:
\begin{eqnarray}
k &=& -\frac{1}{4\pi r^2} \frac{\mathrm{d}}{\mathrm{d}r} 
\int^\infty_{13.6\;\mathrm{eV}/h} \frac{L_\nu e^{-\tau_\nu}}{h\nu}\mathrm{d}\nu, \\
\Gamma &=& -\frac{1}{4\pi r^2} \frac{\mathrm{d}}{\mathrm{d}r} 
\int^\infty_{13.6\;\mathrm{eV}/h} \frac{L_\nu e^{-\tau_\nu}}{h\nu} (h\nu - 13.6~\mathrm{eV}) \;\mathrm{d}\nu.
\end{eqnarray}
We discretize the above equations and take volume averages.

\begin{figure}
	\centering
		\includegraphics[width=7.5cm]{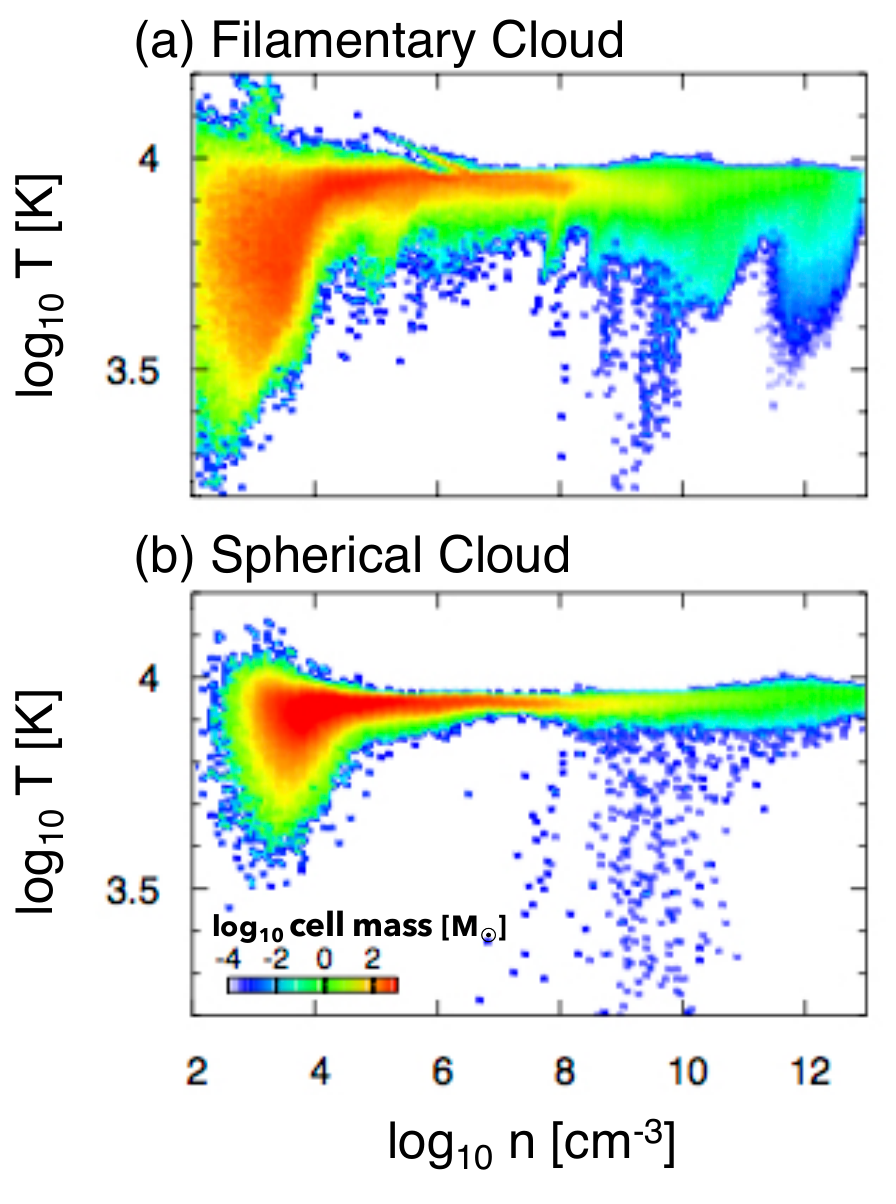}
		\caption{Mass distributions on the density-temperature phase diagram for
                  (a) the filamentary cloud and (b) spherical cloud. 
		 The snapshots are taken when the maximum density exceeds $10^{12}~{\rm cm}^{-3}$ in the early collapse stage. We here divide the whole domain into $200\times200$ cells. The color map represents the gas mass contained within each cell. }
		\label{fig_rhoT_map}
\end{figure}

\begin{figure}
	\centering
		\includegraphics[width=8.5cm]{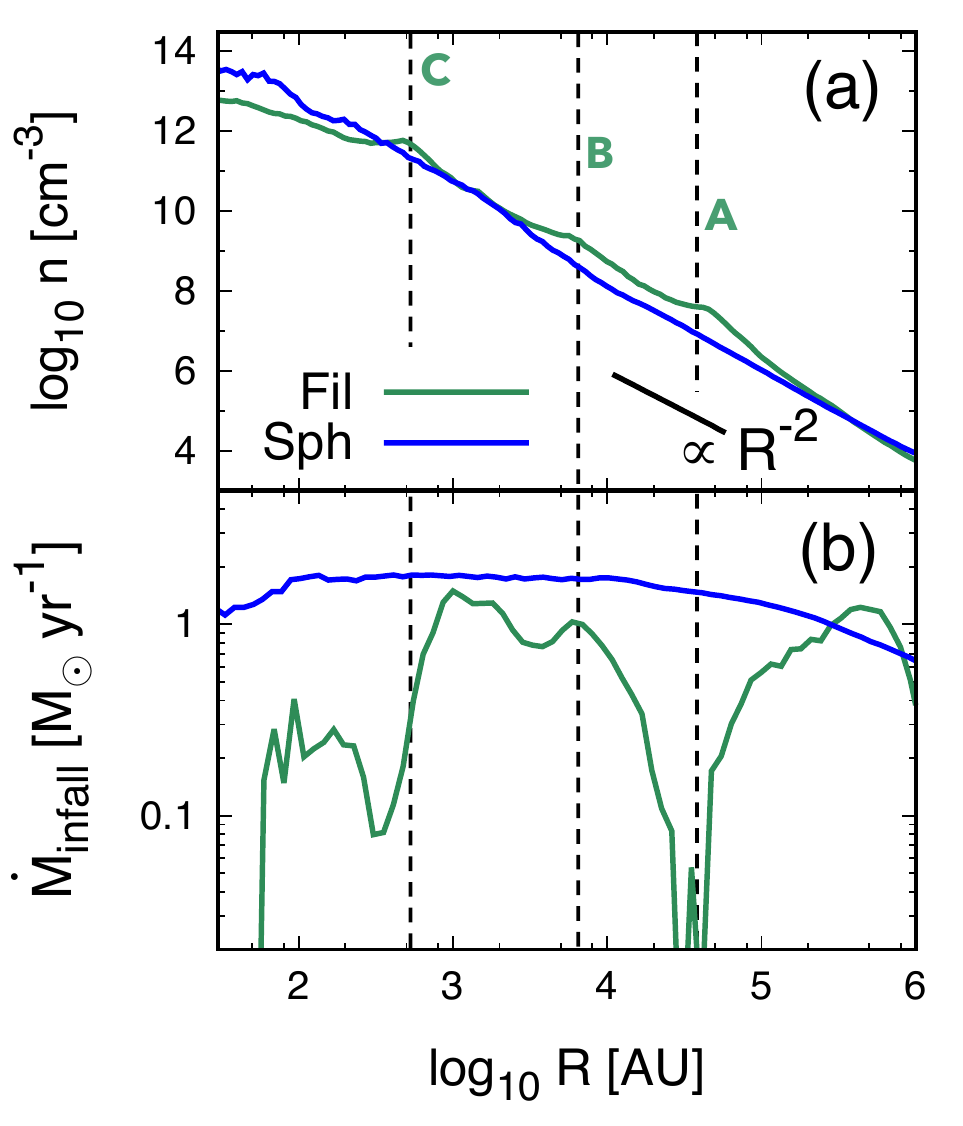}
		\caption{(a) The radial density profiles for the filamentary (green) and the spherical clouds (blue) in their latest stages of the run-away collapse. The vertical dashed lines marked as A, B, and C indicate three characteristic scales where the filamentary cloud shows the density bumps. The power-law profile $n \propto R^{-2}$ is also shown with the short black solid line for the reference. (b) The radial profiles of the mass infall rate ($\dot{M}_\text{infall}$, eq~\ref{eq_Minf}).}
		\label{fig_rhoR_and_fkep}
\end{figure}

\begin{figure}
	\centering
		\includegraphics[width=9.5cm]{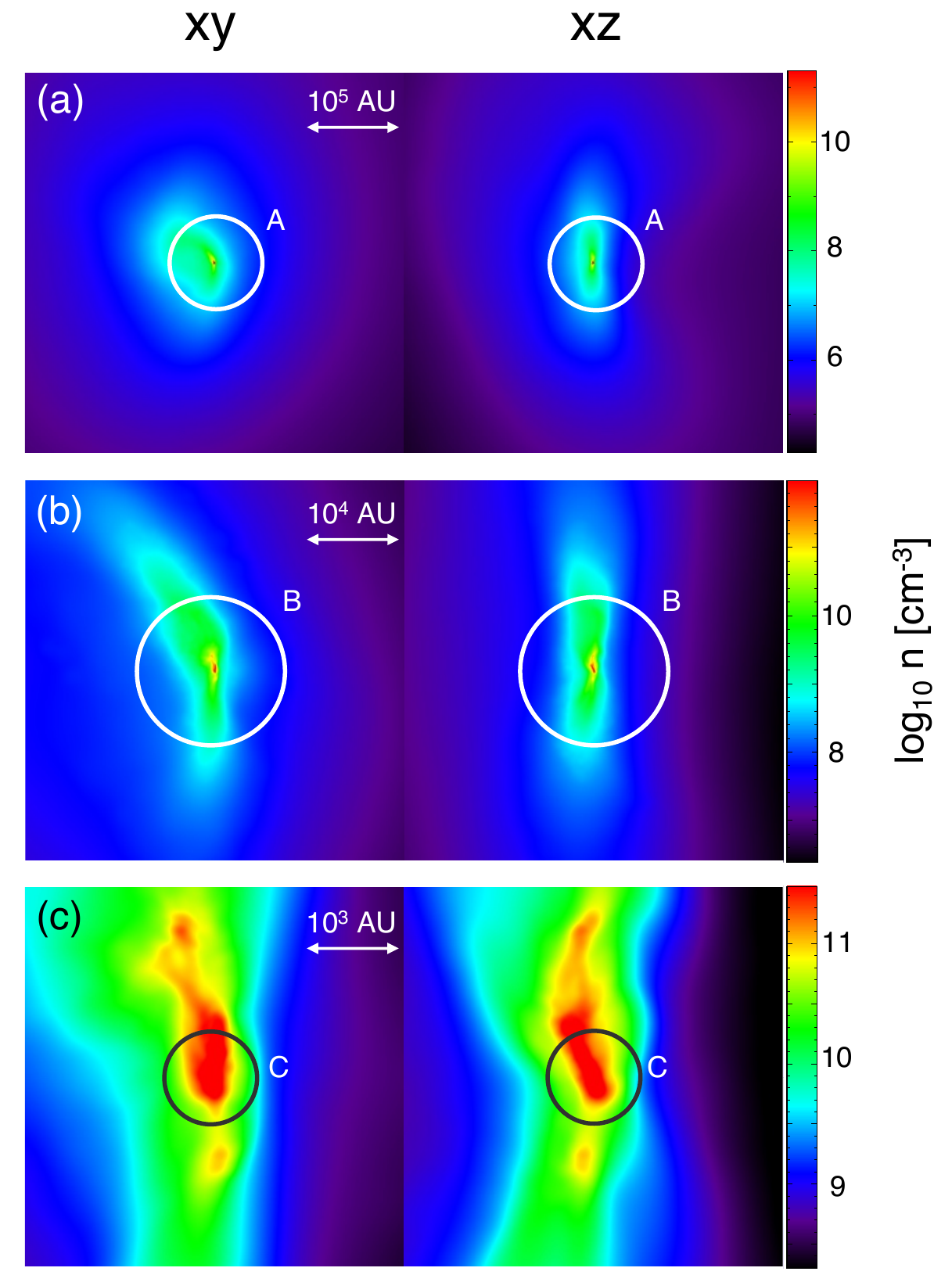}
		\caption{The projected density maps within the filamentary cloud when the maximum density reaches $10^{12}~\mathrm{cm}^{-3}$. The physical scales become smaller from the top to bottom panels.
The left and right panels show the maps projected onto different planes, which are perpendicular to each other (labeled as ``xy'' and ``xz'' planes). The solid circles (A, B, and C) represent the scales presented in Fig.~\ref{fig_rhoR_and_fkep}.}
		\label{fig_density_maps_1100002}
\end{figure}

\begin{figure}
	\centering
		\includegraphics[width=8.8cm]{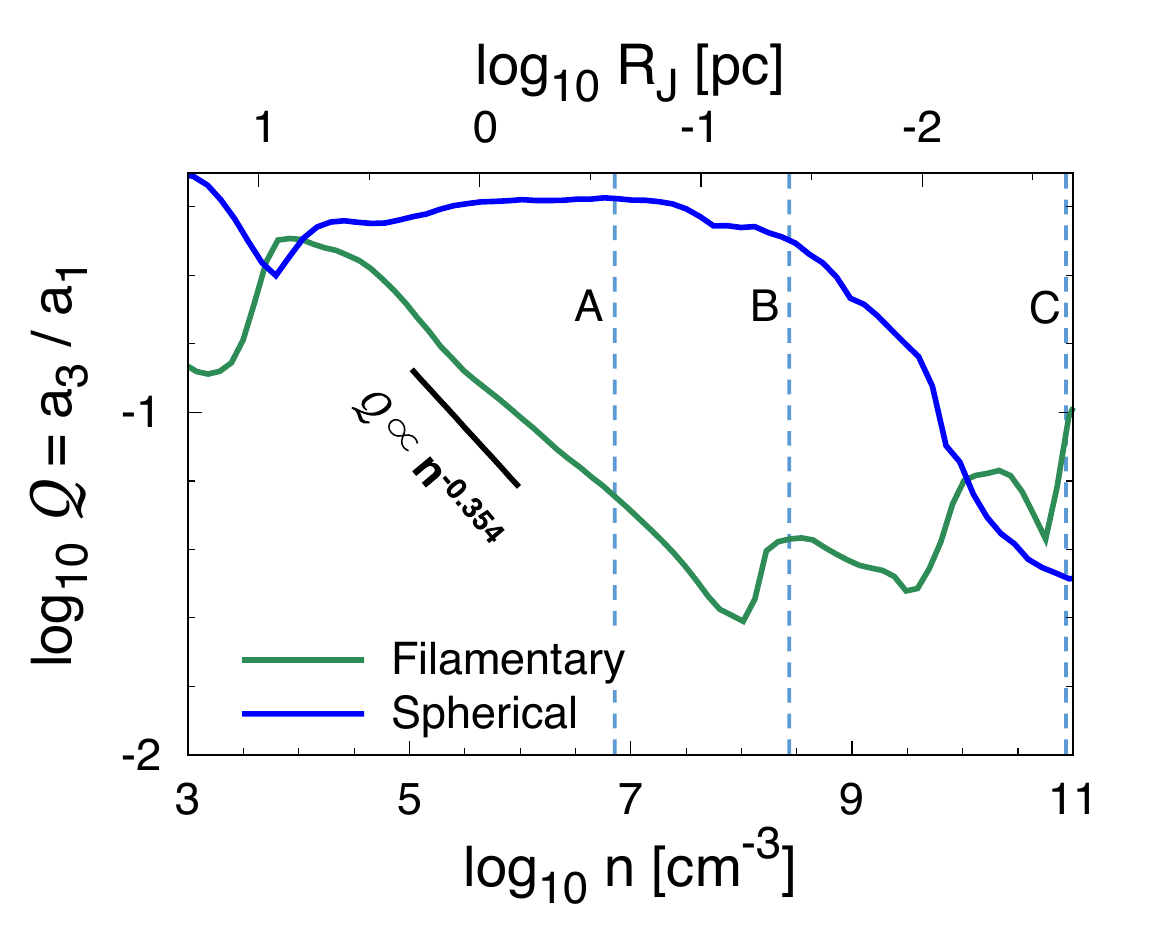}
		\caption{Evolution of the axial ratio $\mathcal{Q}$ during the collapse of the filamentary (blue) and spherical clouds (green). The ratio $\mathcal{Q}$ is defined as $a_{3}/a_{1}$, where $a_1$ and $a_3$ are the lengths of the longest and shortest axes. The upper horizontal axis represents the corresponding Jeans length (eq.~\ref{eq_Jeans_length}) with the temperature fixed at $8000~$K. The vertical dashed lines A, B, and C show the reference scales shown in Fig.~\ref{fig_rhoR_and_fkep}(a).  }
		\label{fig_Axitial_Ratio}
\end{figure}

\section{Results} \label{sec_results}
\subsection{Early collapse phase} \label{sec_early_collapse_phase}
In \cite{Chon+2016}, we followed the cloud evolution until the central density reaches
$10^8~\mathrm{cm}^{-3}$. In both runs, the cloud collapses
in a ``runaway'' manner until a protostellar core appears at the center of the cloud. 
During the collapse, the gas cools by H atomic cooling.
Fig.~\ref{fig_LSS} shows the snapshots when the central density reaches
$10^{12}~\mathrm{cm}^{-3}$. In one case, a filamentary cloud is formed by the strong tidal field from
the nearby galaxy (left top panel in Fig.~\ref{fig_LSS}) whereas
the other cloud contracts nearly spherically (right bottom panel in Fig.~\ref{fig_LSS}).
We call the former as ``filamentary cloud'' and the latter as ``spherical cloud''.

Fig.~\ref{fig_rhoT_map} shows the density-temperature phase diagram within $10~\mathrm{pc}$
from the cloud center. Both clouds collapse almost isothermally on a high-temperature track 
between $8000$ and $10000$~K, 
while the adiabatic expansion 
cools some gas particles slightly below $8000~$K.
In some cases, the adiabatic expansion is followed by H$_2$ formation as 
its collisional dissociation rate decreases with decreasing the temperature.
Especially at $n \gtrsim 10^8~\mathrm{cm}^{-3}$, a small number of gas particles cool
below $5000~$K with the aid of rapid H$_2$ formation by 3-body reaction.
\cite{Inayoshi+2014} point out that the above process triggers the chemical and thermal instability in the early collapse phase.

Fig.~\ref{fig_rhoR_and_fkep}(a) shows the radial density profiles for the filamentary
(green) cloud and for the spherical (blue) cloud when the cloud density reaches $10^{13}~\mathrm{cm}^{-3}$.
In both the cases, the density profile approximately follows $n \propto R^{-2}$ (black),
which is expected for self-similar collapse of an isothermal cloud \citep{Larson1969}.
However, in the filamentary cloud, the profile shows small bumps.
We mark them in Fig.~\ref{fig_rhoR_and_fkep}(a) by the dashed lines and
label them as A, B, and C from low to high density. Fig.~\ref{fig_rhoR_and_fkep}(b)
shows the radial profiles of the mass infall rate ($\dot{M}_\text{infall}$) measured at the distance $R$,
\begin{equation} \label{eq_Minf}
\dot{M}_\text{infall} (R) \equiv 4\pi R^2 \rho v_\text{rad},
\end{equation}
where $\rho$ is the density and $v_\text{rad}$ is the radial velocity of gas.
We find that the spherical cloud has $\dot{M}_\text{infall} \gtrsim 1~M_\odot~\mathrm{yr}^{-1}$
at $R \lesssim 10^6~$AU. The filamentary cloud also has $\dot{M}_\text{infall} \gtrsim 1~M_\odot~\mathrm{yr}^{-1}$ at  $R \lesssim 10^6~$AU. However, around the indicated regions A and C,
$\dot{M}_\text{infall}$ is smaller than $1~M_\odot~\mathrm{yr}^{-1}$.
As we will see later in Section \ref{sec_sink_evo_and_fragmentation},
the filament starts to fragment at A and C in the later accretion phase. 

Fig.~\ref{fig_density_maps_1100002} shows the projected density distribution
around the protostar in the filamentary cloud. The left and right panels show
the projections from the different directions and we label them as ``xy'' (left) and ``xz'' (right).
In the region larger than A, the cloud has round shape projected onto ``xy'' plane
while it is elongated and shows a filamentary structure in the ``xz'' plane.
Near region B, the cloud shows a clear filamentary structure.
Around scale C, the filament starts to fragment into multiple clumps.
We will see the cloud also fragments at scale A at 6800~yr after the
central protostar formation (Section~\ref{sec_sink_evo_and_fragmentation}).

To quantify the deviations from spherical collapse,
we approximate iso-density volumes around the cloud center by a series of ellipsoids.
In Fig.~\ref{fig_Axitial_Ratio}, we show how the axial ratio $\mathcal{Q} \equiv a_3/a_1$
of the iso-density contour varies with different densities for the filamentary (blue)
and the spherical (green) clouds. Here, $a_1$ and $a_3$ are the lengths of the axes
which have the largest and smallest magnitudes, respectively. The filamentary cloud
has smaller axial ratios than the spherical cloud. For example, at $n = 10^8~\mathrm{cm}^{-3}$,
the filamentary and spherical clouds have the axial ratios of $\sim 0.03$ and $\sim0.3$, respectively.
In Fig.~\ref{fig_Axitial_Ratio}, we also show the Jeans length ($R_\text{J}$) as a
function of the gas density ($n$) with the fixed temperature $T=8000~\mathrm{K}$,
\begin{align} \label{eq_Jeans_length}
R_\text{J} = 0.22~\mathrm{pc} \left( \frac{n}{10^{7}~\mathrm{cm}^{-3}} \right )^{-1/2}.
\end{align}

The axial ratio decreases at $n \gtrsim10^{10}~\mathrm{cm}^{-3}$ in the spherical cloud
because the disk-like structure appears around owing to
the finite angular momentum. Contrastingly, the small axial ratio at $n \gtrsim 10^6 ~\mathrm{cm}^{-3}$
of the filamentary cloud is not caused by disk formation.
The corresponding Jeans scale at this density is much larger
than the scale where the angular momentum support is strong
at this epoch. Rather, the small axial ratio is explained by development
of a bar in isothermal gas. \cite{Hanawa+2000} suggest a relation $\mathcal{Q}\propto n^{-0.354}$
in linear regime. We see that $\mathcal{Q}$ actually follows this relation
in the filamentary cloud (black solid line). At $n\sim 10^8~\mathrm{cm}^{-3}$, $\mathcal{Q}$
reaches 0.03 and enters the nonlinear regime \citep{Tsuribe+2006,Chiaki+2016}. 

Note that the elongation, i.e., the decrease of $\mathcal{Q}$, is not entirely driven by the tidal torque
from the nearby massive galaxy. The tidal radius is about 1 pc and thus
has almost no effect at smaller length scales.
The initial perturbations are seeded by the tidal torque, and grow through isothermal collapse
in the filamentary cloud.

\begin{figure}
	\centering
		\includegraphics[width=6.5cm]{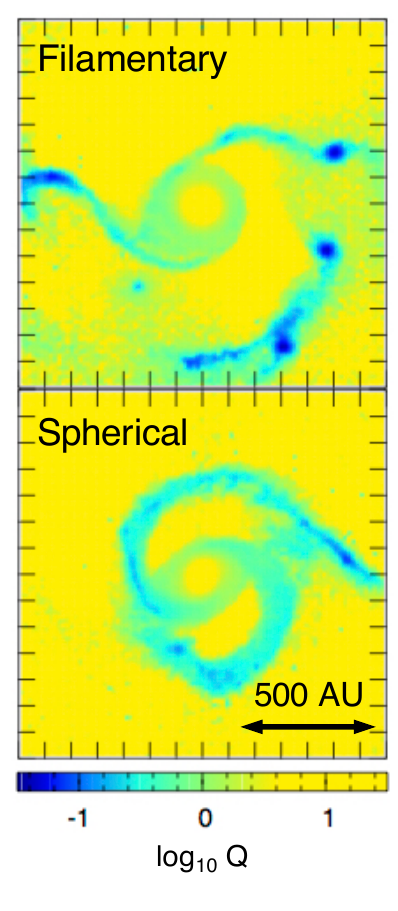}
		\caption{Distributions of Toomre $Q$ parameter (eq.~\ref{eq_ToomreQ}) around the central protostar for the filamentary (top) and the spherical (bottom) clouds. The snapshots are taken at $\simeq 1500~$yr after the central protostar is formed. }
		\label{fig_Toomre_Q}
\end{figure}

\begin{figure*}
	\centering
		\includegraphics[width=16.cm]{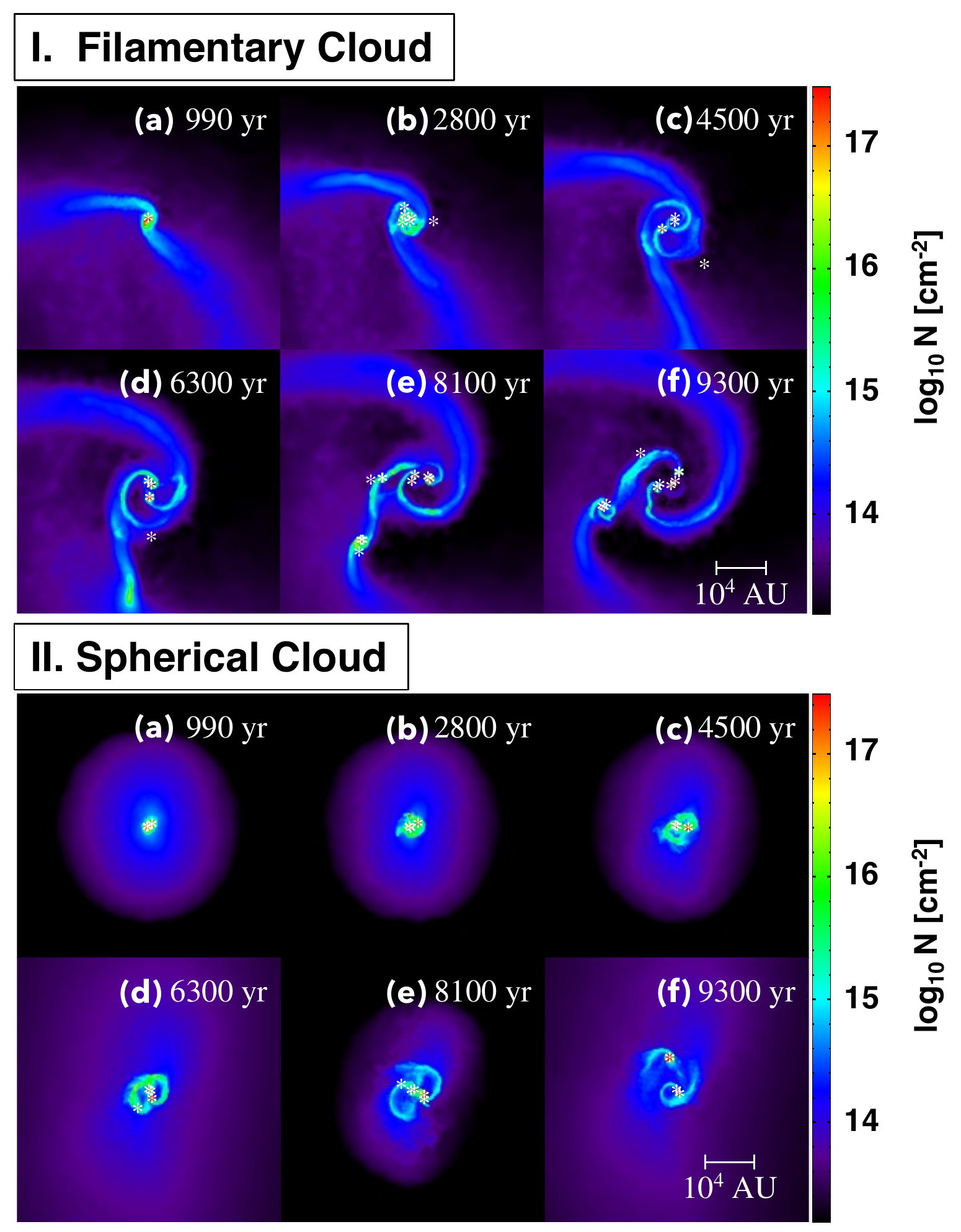}
		\caption{Face on view of the evolving disk and stellar systems in the filamentary (top) and spherical (bottom) clouds. The white asterisks and colormaps represent the protostars and column density of the surrounding gas. The time is measured from the epoch when a protostar first appears in the clouds.} 
		\label{fig_disc_pics}
\end{figure*}

\begin{figure*}
	\centering
		\includegraphics[width=16.5cm]{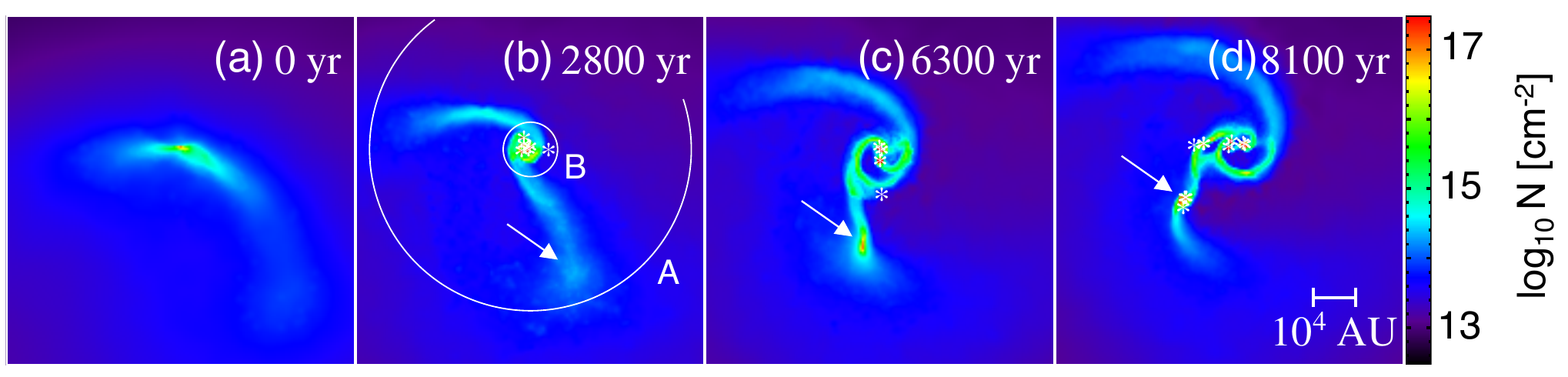}
		\caption{Filament fragmentation occurring in the filamentary cloud. The arrows indicate where a star-disk system forms through such an event. In panel (b), the white circles indicate the different spatial scales A and B appeared in Fig.~\ref{fig_rhoR_and_fkep}. }
		\label{fig_filament_fragmentation}
\end{figure*}

\begin{figure*}
	\centering
		\includegraphics[width=15.5cm]{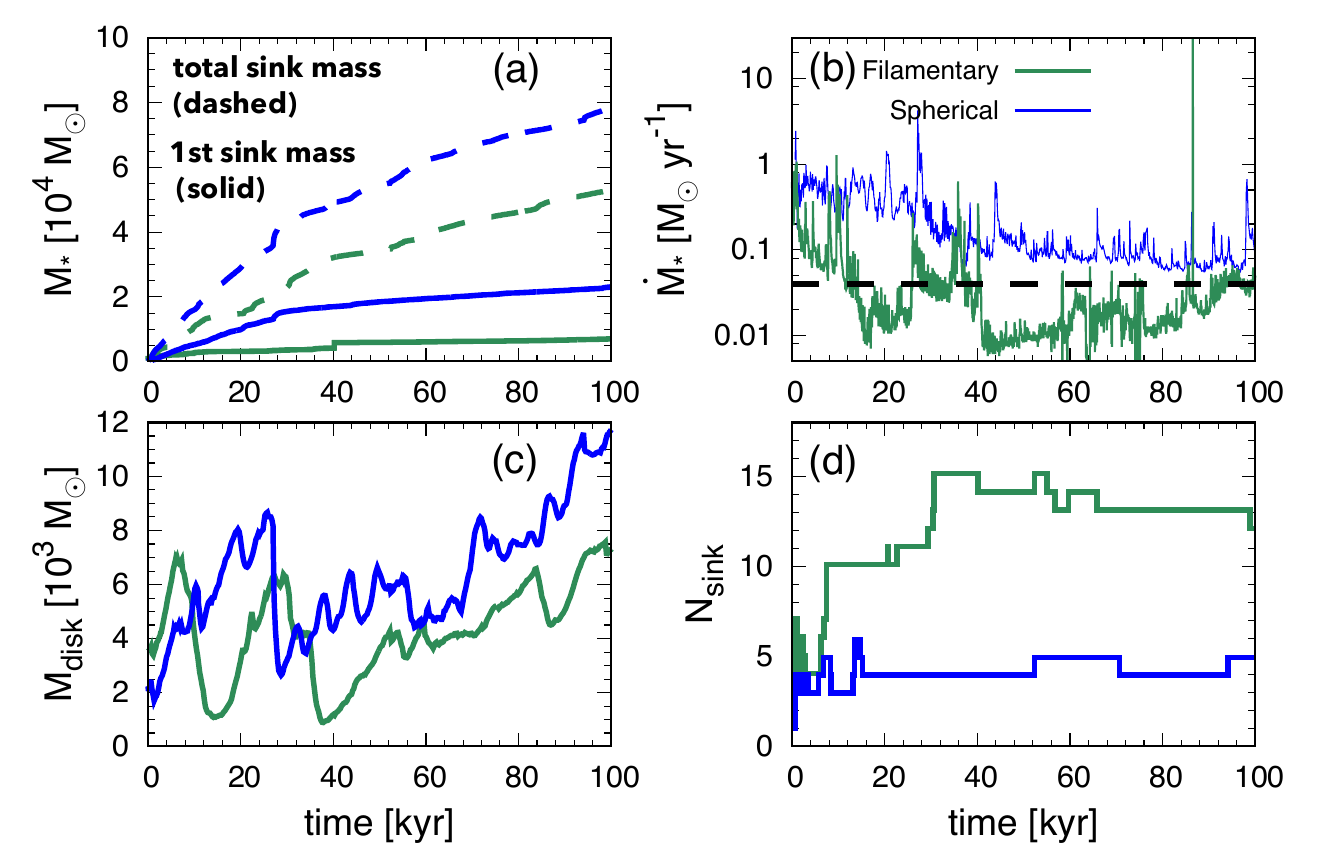}
		\caption{Evolutions of (a) the stellar mass, (b) accretion rate, (c) disk mass, and (d) number of the protostars for the filamentary (green) and spherical (blue) clouds. In panel (a), the solid and dashed lines represent the masses of the primary protostars and total stellar masses. The primary protostar is the one which is first formed in the calculation.
In panel (b), the horizontal dashed line shows the critical accretion rate, below which the star contracts to the ZAMS phase (see Section~\ref{sec_SEM}). The accretion rate is averaged over $30~$yr.
		} 
		\label{fig_sink_masses}
\end{figure*}

\begin{figure*}
	\centering
		\includegraphics[width=13.5cm]{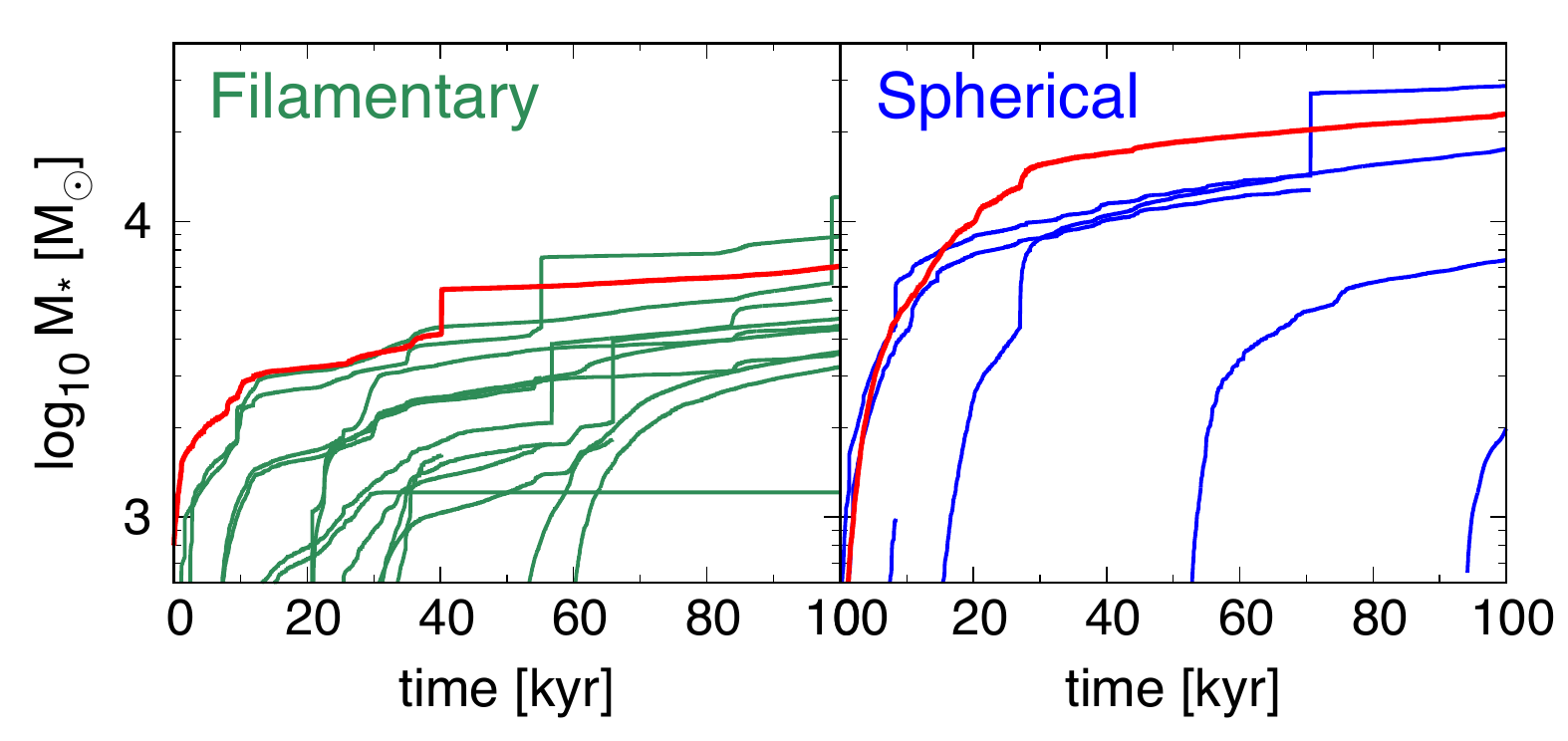}
		\caption{Mass evolution of all the protostars in the filamentary (left)
                  and the spherical (right) clouds. 
		The stellar merger events are marked with the vertical jumps seen in some lines. Note that another line ends at the same point, representing the merger partner.
		 The red line shows
		the mass evolution of the primary protostar.} 
		\label{fig_sink_mass_evolution_all}
\end{figure*}

\begin{figure}
	\centering
		\includegraphics[width=7.5cm]{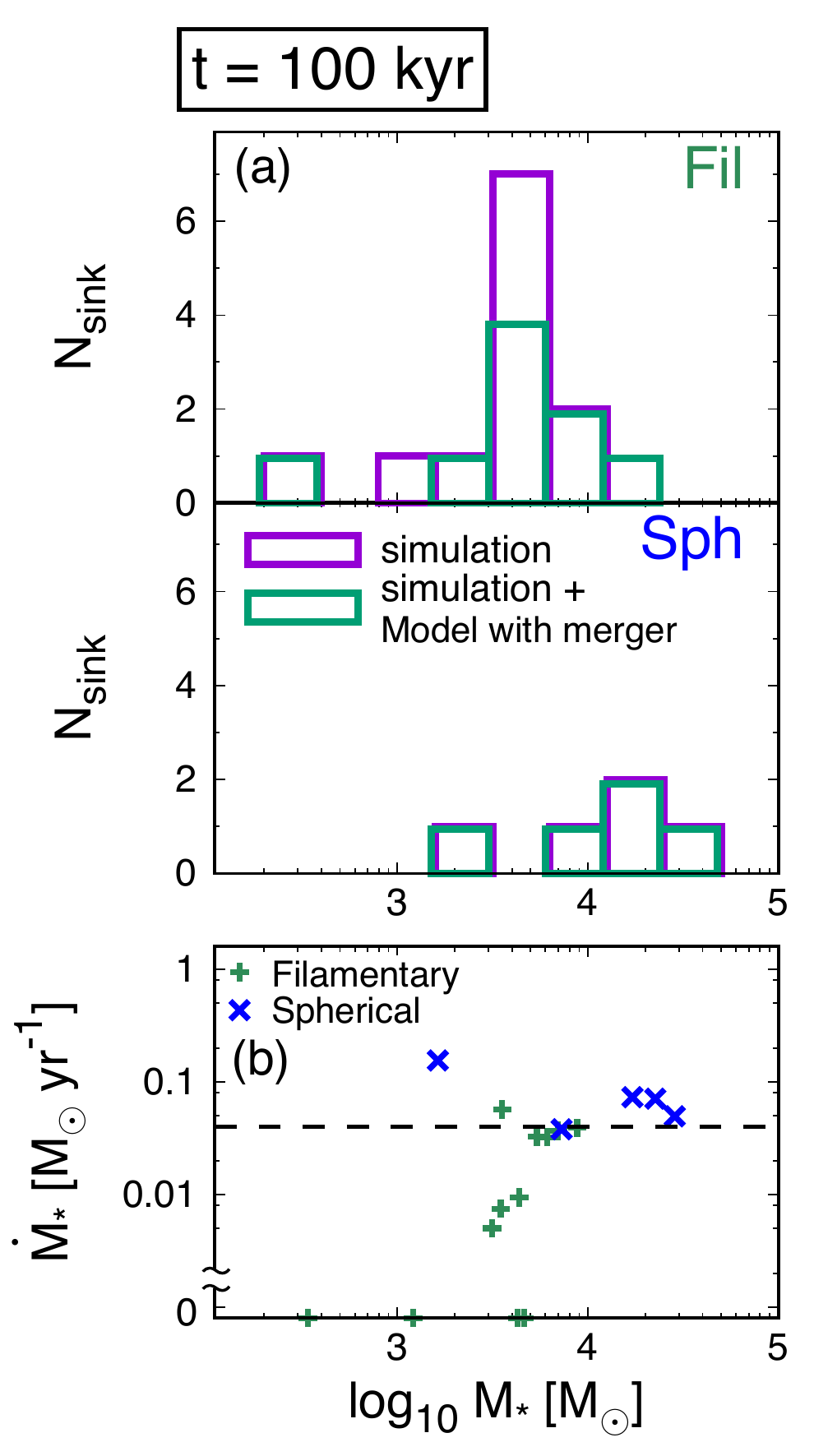}
		\caption{(a) Stellar mass distributions at $t=100~$kyr for the filamentary (top) and spherical (bottom) clouds. 
In each panel, the purple histograms show the simulation results, and the green ones show the modified distributions with ``Model with merger''.
		In this model, we first pick up protostar pairs which experience close encounters with $< 200~$AU and assume these pairs have merged into single protostars.
The mass of a combined protostar is the total mass of the original pair.
		(b) Mass accretion rates ($\dot{M}$) against the stellar masses ($M_*$) 
		at $t=100~$kyr for the filamentary (green) and the spherical (blue) clouds.
		The dashed line represents the critical mass accretion rate, 
		below which the protostar contracts into the ZAMS phase.
		} 
		\label{fig_mass_func}
\end{figure}

\subsection{Later accretion phase}
After the central protostar is formed, the infalling gas with finite angular momentum is accumulated 
to form a circumstellar disk. 
In the self-gravitating gas disk, angular momentum is transferred outward by the gravitational
torque and the central protostar is able to accrete and grow in mass.
When the disk becomes massive enough, however, it bears multiple fragments.
In this section, we study the evolution of the protostars and the circumstellar disks.

\subsubsection{The evolution of the star-disk system} \label{sec_disk_sink_evo}
A heavy circumstellar disk quickly undergoes gravitational fragmentation.
To quantify the gravitational instability,
we calculate the Toomre-$Q$ parameter defined by
\begin{equation} \label{eq_ToomreQ}
Q = \frac{c_\text{s} \Omega}{\pi G \Sigma},
\end{equation}
where $c_\text{s}$ is the sound speed, $\Omega$ is the orbital frequency, and $\Sigma$ is the surface density of the disk. 
Fig.~\ref{fig_Toomre_Q} shows the distribution of $Q$ parameter
on the disk plane at $1500~$yr after the protostar is formed. 
Inside the disk, $Q$ is order of unity while it is smaller than unity along the spiral arms \citep{Gammie2001, Takahashi+2016}. 
These arms quickly fragment into multiple clumps, where the $Q$ is smaller than unity ($Q \sim 0.1$).

The most unstable length- and mass-scales for the gravitational fragmentation are given
by $\lambda \sim c_\text{s} / \Omega$ and $M_\text{frag} \sim \pi(\lambda/2)^2\Sigma$
where $Q = 1$ has been assumed.
With the typical temperature and surface density in our simulations, these scales are 
\begin{align}
\lambda  = \frac{\pi c_\text{s}^2}{G\Sigma} 
&= 17~\mathrm{AU} \left ( \frac{T}{8000~\mathrm{K}} \right ) \left ( \frac{10^5~\mathrm{g ~cm^{-2}}}{\Sigma} \right ), \\
M_\text{frag} = \frac{\pi^3 c_\text{s}^4}{4G^2 \Sigma}
&= 2.5~M_\odot \left ( \frac{T}{8000~\mathrm{K}} \right )^2 \left ( \frac{10^5~\mathrm{g ~cm^{-2}}}{\Sigma} \right )^{2}.
\end{align}
We find that the initial fragment mass is $\sim 5~M_\odot$, in good agreement with
the above simple estimation.

Fig.~\ref{fig_disc_pics} shows the development of multiple protostar-disk systems
for the filamentary and spherical cases, where the color scales indicate the column density
distribution projected onto the disk plane.
The circumstellar disk continuously acquires mass and grows in mass and size
(Fig.~\ref{fig_disc_pics}-Ia and IIa).
Then spiral arms are excited, which further fragment into multiple clumps
that finally contract to become protostars (Fig.~\ref{fig_disc_pics}-Ib and IIb).
At early epochs, some of the newly formed protostars
rapidly fall onto the central protostar. The inward migration 
is promoted by the interaction with the gas inside the disk (the so-called
Type-I migration).

The gas density near the disk center decreases with time because remaining protostars
accrete the surrounding gas (Fig.~\ref{fig_disc_pics}-Ic and IId). 
This makes the gas-star interaction less efficient
and the migration toward the central protostar ceases. 
More than three protostars are formed inside the disk and they remain
on stable orbits for several tens of the disk rotation period.
These features are common in both the filamentary and the spherical clouds. 

In the spherical cloud, only a single circumstellar disk is formed around the cloud center.
The multiple stellar system formed via fragmentation of the primary disk
is stable, and no stars are found to be ejected from the disk
until the end of our calculation ($t\sim 0.1~$Myr). 
The separations between the protostars are smaller than the disk size
of $\lesssim 10^4~$AU at $0.1~$Myr.

In the filamentary cloud, another star-disk system is formed at $\sim 10^4$ AU
from the cloud center as seen in Fig.~\ref{fig_disc_pics}-Ie.
In panel (b), we indicate the radii corresponding to A and B in Fig.~\ref{fig_rhoR_and_fkep} by white circles.
The filament starts to fragment just near the circle A,
where the filament's axial ratio is
small enough and an overdense clump can grow in a non-linear fashion
(see Section~\ref{sec_early_collapse_phase}).
The clump soon migrates toward the cloud center (panel d). 
Since angular momentum is difficult to be extracted during its migration due to the low gas density,
the final separation between the two star-disk systems is set
by the angular momentum barrier.
The separation is larger than that found in the spherical cloud.
The filamentary collapse tends to generate multiple star-disk systems
that are well-separated.

\subsubsection{Stellar mass growth} \label{sec_sink_evo_and_fragmentation}
Fig.~\ref{fig_sink_masses}(a) shows the evolution of the total protostellar mass (dashed)
and the mass of the primary protostar (solid line).
Fig.~\ref{fig_sink_masses}(b) shows the evolution of the mass accretion rate
onto the primary protostar.
Fig.~\ref{fig_sink_mass_evolution_all} shows the mass evolution of all
the protostars found in each simulation. In these figures, the green and blue lines represent
the results for the filamentary and spherical clouds, respectively. 

The accretion rates are highly variable with time.
Such rapid variations have been often seen for the mass accretion
through a self-gravitating circumstellar disk, and
is known as ``episodic accretion'' \citep[e.g.][]{Vorobyov+2013}.
We notice that the accretion rate suddenly increases up
to $\sim 1$--$10~M_\odot~\mathrm{yr}^{-1}$ at $t\sim 27~$kyr
in the filamentary cloud. At this epoch, the primary protostar
experiences the three body interaction with other protostars.
The close encounter of the primary protostar with another one
excites strong density enhancement and causes
a large amount of the gas to fall onto the primary protostar. 

In the spherical cloud, stars grow in mass with a rate of $0.1 - 1~M_\odot~\mathrm{yr}^{-1}$
until the end of the calculation.
The rates are much higher than the critical accretion rate,
$\dot{M}_\text{crit} = 0.04~M_\odot~\mathrm{yr^{-1}}$,
and thus the stellar envelope remains expanded to the radius
of several $\times 10$--$100~$AU.
In the spherical gas cloud, the ionizing photon emissivity
is smaller than the number of hydrogen atoms infalling toward the center
per unit time \citep{Sakurai+2015},
and hence the protostars
cannot fully ionize the surrounding gas.

In the filamentary cloud, however, $\dot{M}_*$ decreases occasionally
below $\dot{M}_\text{crit}$. 
Fig.~\ref{fig_sink_masses}(c) shows the evolution of the disk mass,
where the disk mass is defined as the gas mass with
$n > 10^9~\mathrm{cm}^{-3}$.
At $t\sim 10~$kyr, another protostar is formed in the disk,
and both $\dot{M}_*$ and the disk mass rapidly decrease
because the newly born protostar accretes the gas in its surrounding.
The primary protostar can then contract toward the ZAMS under the slow accretion,
and the emissivity of ionizing photons increases.
Nevertheless, the ionizing radiation does not prevent the mass accretion
because the surrounding gas, even if it is ionized,
is strongly bound by the protostar. 
By $t \sim 25~$kyr, the gas is quickly accumulated around the protostar
and the accretion rate rises over the critical value.
The star recovers the supergiant phase again.
We will discuss the efficiency of the ionizing radiation feedback
later in Section~\ref{sec_radiation_feedback}.

Fig.~\ref{fig_sink_masses}(d) shows the number of the protostars found in the simulation.
The disks around the protostars become more stable with time in both clouds.
In fact, the total stellar mass grows at a faster rate than the disk mass.
Thus, the Toomre-$Q$ becomes larger with increasing time (see eq.~\ref{eq_ToomreQ}).
The number of protostars hardly changes after $t\sim 40$ and $20~$kyr for the filamentary
and the spherical clouds, respectively. In the filamentary cloud, fragmentation still continues
at later time than in the spherical cloud. This is because the fragmentation also occurs at the filament far from the disk (see Figs.~\ref{fig_disc_pics}-Id, Ie, and \ref{fig_filament_fragmentation}). The filament fragmentation itself is not stabilized by the growth of the protostars. For example, filament fragmentations take place at $t\sim 8$ and $30~$kyr. We can see that the number of protostars suddenly increases around these epochs.

\subsubsection{Mass distribution of the protostars}
Fig.~\ref{fig_mass_func}(a) shows the mass distributions of the protostars at $t = 100~$kyr for the filamentary (top) and the spherical (bottom) clouds. The purple histograms show the mass distribution obtained in the simulations.
Since the resolution of our simulation is limited, further migration
within $\sim 100~$AU cannot be followed accurately (Section~\ref{sec:resolution}). 
Since the stellar radius is also of the order of $\sim 100~$AU,
the two protostars are likely to merge,
and the stellar mass of each star obtained in our simulation is probably underestimated. 
Once a protostar pair gets closer than $200~$AU,
they are assumed to merge. Thereafter we treat the pair as a single protostar,
and just sum the masses of the merging protostar pair. We refer this model as ``Model with merger''
(green histograms in Fig.~\ref{fig_mass_func}). 

For the filamentary cloud, the stellar mass distribution is concentrated
at $3000$--$6000~M_\odot$. Fig.~\ref{fig_mass_func}(b) summarizes how rapid mass accretion still continues depending 
at the epoch of $t=100~$kyr. The dashed line represents the critical mass accretion rate below which the supergiant star contracts into the main sequence phase. 
In the spherical cloud, protostars grow at rates $\gtrsim 0.1~M_\odot~\mathrm{yr}^{-1}$. Even the most massive one remains in the supergiant phase.
Consequently, the protostars cause only weak ionizing radiation feedback for entire period of our calculation. 
In the filamentary cloud, the accretion rate decreases with time and becomes smaller than the critical value for most
of the protostars at $t=100~$kyr. The total ionizing photon emissivity increases to cause strong radiation feedback.
However, as we will see in Section~\ref{sec_radiation_feedback}, the ionizing radiation does not completely prevent mass accretion
in our calculation, because the accreting gas is strongly bound by the central stars.

\begin{figure}
	\centering
		\includegraphics[width=8.0cm]{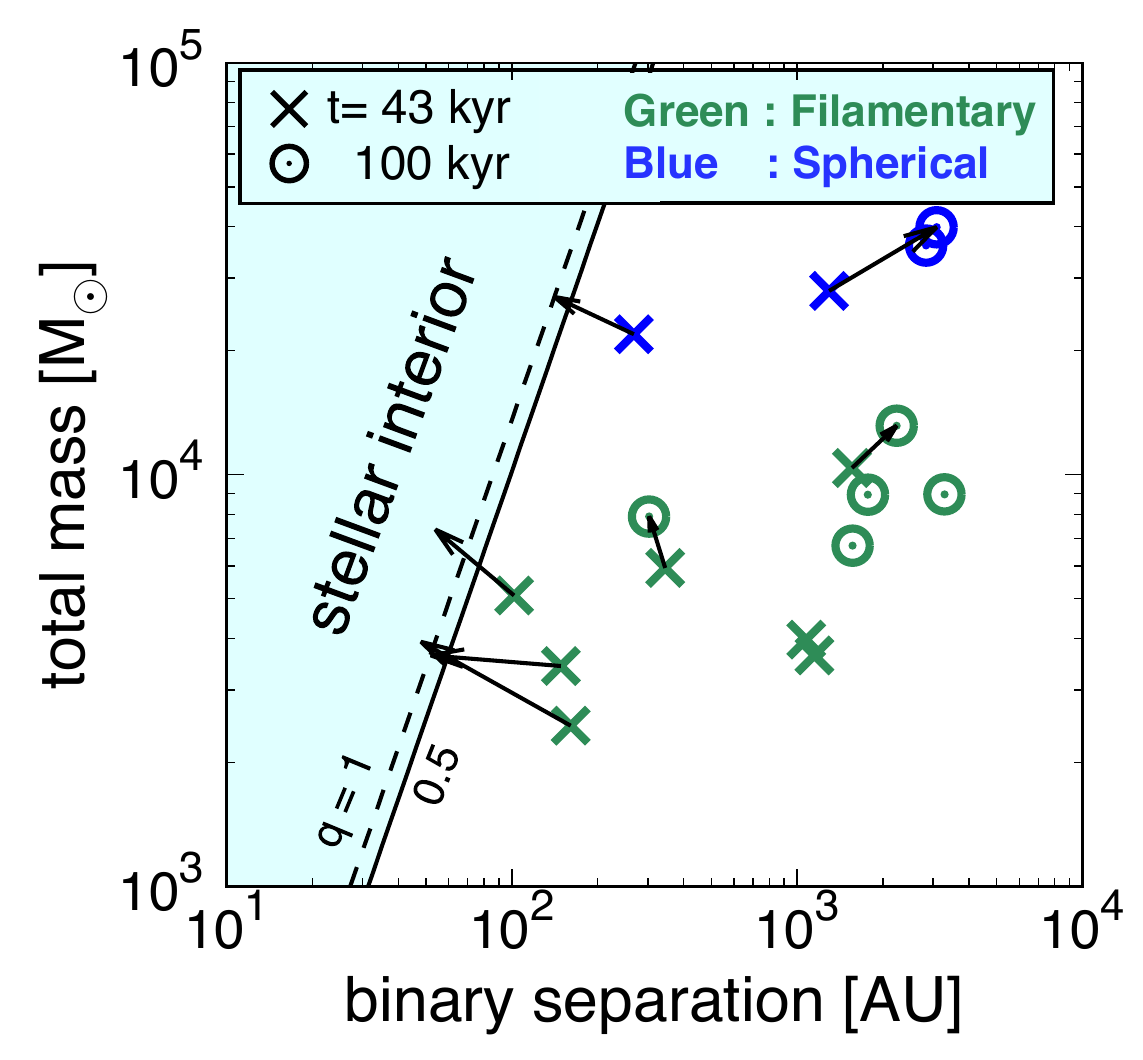}
		\caption{Appearance and evolution of the stellar binaries in the filamentary (green) and spherical (blue) clouds. The different symbols represent the different epochs $t = 43~$kyr (crosses) and $100~$kyr (circles). The solid and dashed lines show the radii of the binary stars, assuming they are in the supergiant phase (eq.~\ref{eq_rstar}) with the different mass ratios of $q=0.5$ and 1. Once a binary enters the shaded region, it is assumed to be merged in our simulations. The arrows indicate the evolution of the same binaries between the two presented epochs.
		} 
		\label{fig_binary_props}
\end{figure}

\begin{figure}
	\centering
		\includegraphics[width=9.cm]{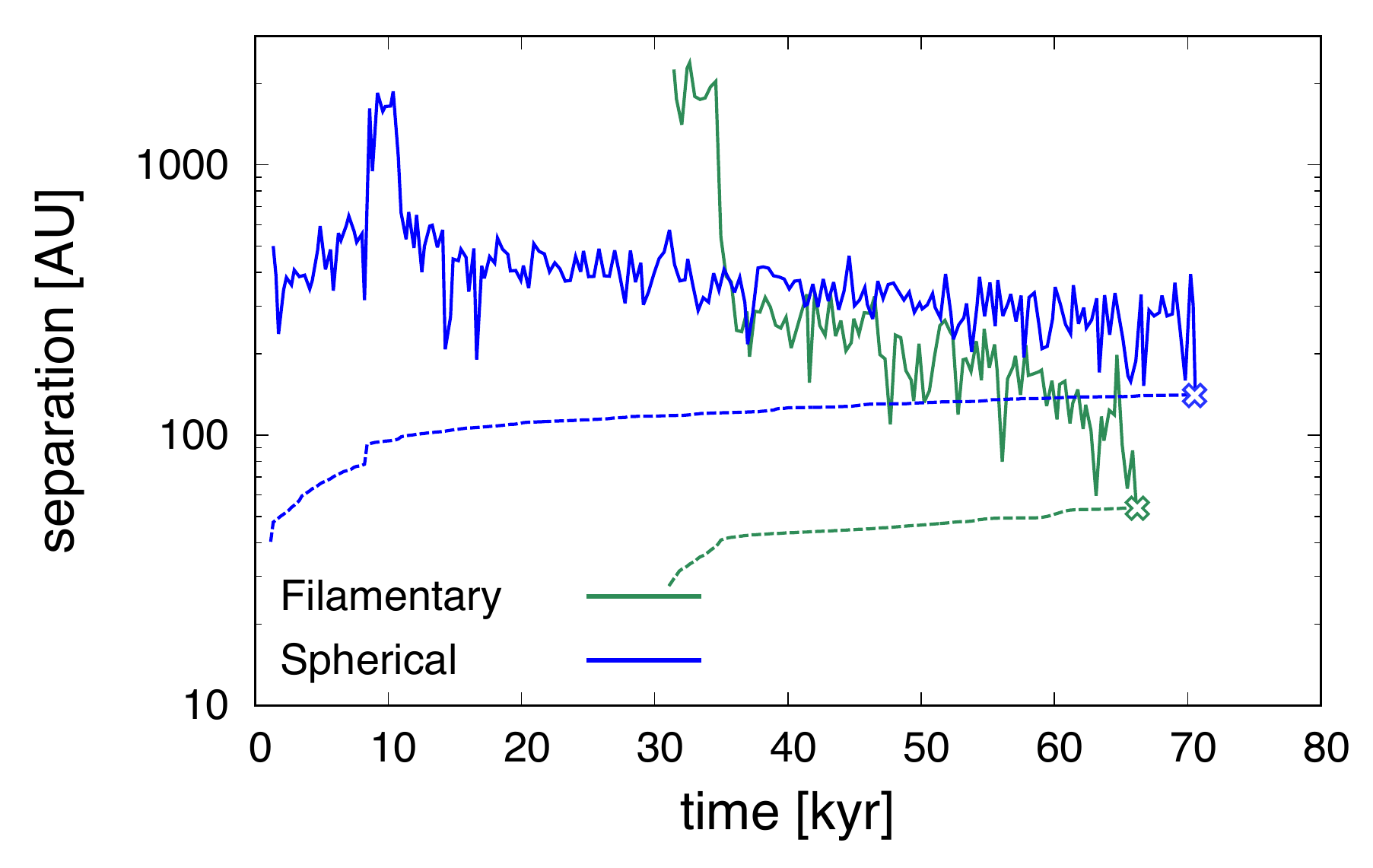}
		\caption{Examples of the time evolution of the binary separation. Shown are merged binaries typical in the filamentary (green) and spherical (blue) clouds. The solid and dashed lines represent the binary separation and the radii of the binary stars. The start and end points of each line correspond to the epochs of the binary formation and merger.} 
		\label{fig_binary_separation}
\end{figure}

\begin{table}
	\centering
	\scalebox{1.}[1.]{
	\begin{tabular}{ccccc} \\[2pt] \hline \hline  
	       ($t=100~$kyr)                   & total & survived & ejected & binary \\[2pt] \hline \\[0.25pt] 
	Filamentary & 25   & 13 & 3 & 7 \\[2pt]
	Spherical     & 13   & 5   & 0 & 4 \\[2pt] \hline \hline
	\end{tabular}
	}
	\caption{Statistical properties of protostars at $t=100~$kyr. 
	Each column show the number of protostars in total, survived, ejected, and belong to any binaries at $t=100~$kyr
	from left to right. }
	\label{tab_sink_pairs}
\end{table}

\subsubsection{Merger and ejection of protostars} \label{sec:merger_ejection}
In our simulations, protostars interact gravitationally with other protostars.
The close encounter of protostars results in various events such as binary formation, merger, and ejection of the protostars. 
Mergers mainly take place within $\lesssim 1$--$2~$kyr after the disk fragmentation because the surface density of the disk decreases
and the disk-star interaction becomes inefficient after the sequence of fragmentation.
Table~\ref{tab_sink_pairs} summarizes the number of protostars which are formed in the simulation. We also list the number of
survived, ejected, and belong to any binaries at $t=100~$kyr for the filamentary and the spherical clouds.
A protostar is assumed to be ejected once the velocity exceeds the escape velocity ($v_\text{esc}$) of the cloud,
\begin{equation}
v_\text{esc} = \sqrt{\frac{2GM_\text{enc}(<R)}{R}},
\end{equation}
where $R$ is the distance from the center of mass. Once a protostar pair becomes gravitationally bound, the pair is classified as a binary.
We judge a pair is bound if the total energy ($E_\text{tot}$) 
\begin{equation}
E_\text{tot} = \frac{m_1 v_1^2}{2} + \frac{m_2 v_2^2}{2} - \frac{Gm_1m_2}{r}
\end{equation}
is negative, where $m_i$ is the mass of the protostar member, $v_i$ is the relative velocity to the center of mass, and $r$ is the separation of the members ($i=1,2$).

\begin{figure*}
	\centering
		\includegraphics[width=15.5cm]{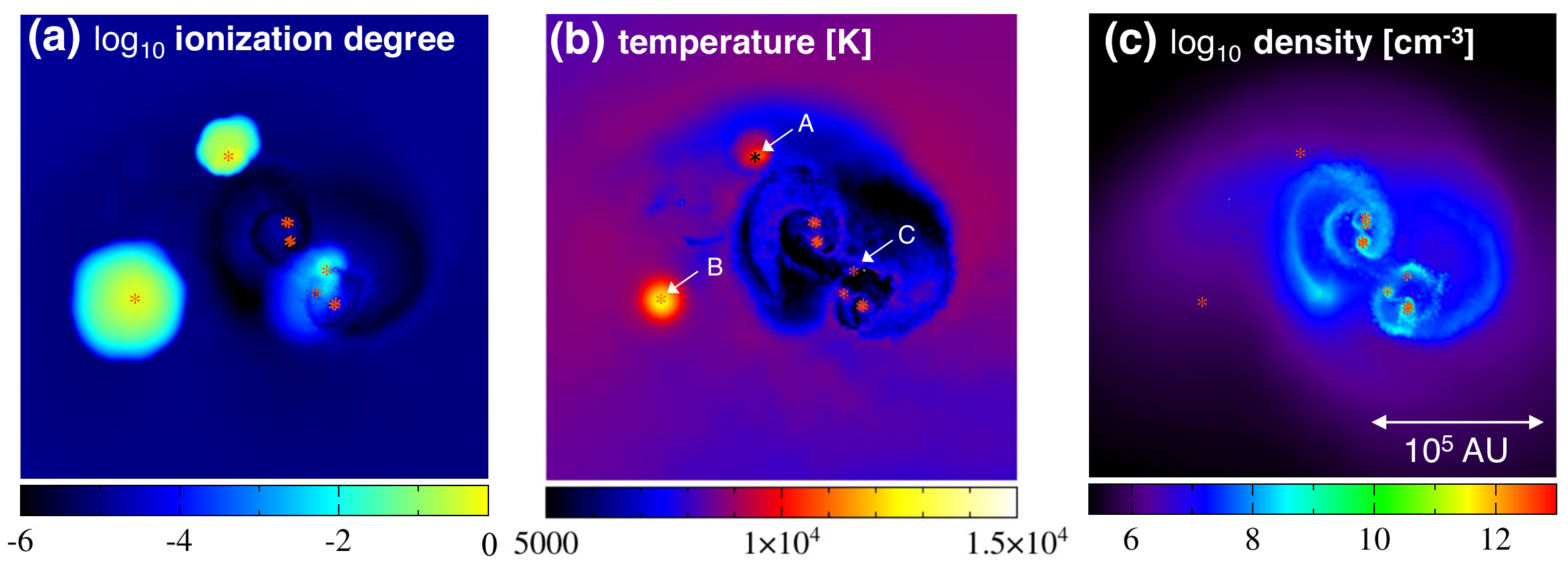}
		\caption{Spatial distributions of (a) hydrogen ionization degree, (b) temperature, and (c) density in the filamentary cloud at $t=38~$kyr. The asterisks mark the positions of the protostars. In panel (b), protostars which emit a copious amount of ionizing photons are marked with the white arrows and labeled as A, B, and C.
		} 
		\label{fig_1100002_Xion_maps}
\end{figure*}

Fig.~\ref{fig_binary_props} shows the distributions of separations against the total mass of binaries at different epochs
of $t=43$ (cross) and $100~$kyr (open circle). We present the evolution of the mass and the separation of the binaries by arrows.
The solid and dashed lines represent the stellar radius under the mass ratios ($q$) with $1$ and $0.5$, respectively.
Inside the shaded region, the separation is smaller than the stellar radius. We confirm the binaries usually have $q \gtrsim 0.5$.
We also assume both of the member stars are accreting material at a rate of $>0.04~M_\odot~\mathrm{yr}^{-1}$ and stars are in the supergiant phase (eq.~\ref{eq_rstar}). 
At $t=43~$kyr, some binaries have separations slightly larger than the stellar radius. These stars grow in mass and radius
with time (eq.~\ref{eq_rstar}) and finally coalesce.  At $t=100~$kyr, the survived binaries have typical separations
of a few times $\sim 10^3~$AU, which is an order of magnitude
larger than the stellar radius. These binaries are so stable that they will finally evolve into BH binaries after the stellar
lifetime. We will discuss the evolution of these binaries in Section~\ref{sec_discussion_binary_evo}.

Fig.~\ref{fig_binary_separation} shows the evolution of the separation of the merged binaries. Here, we show examples both of the filamentary (green) and of the spherical (blue) clouds. 
The solid and dashed lines show the evolutions of binary separation and the stellar radius, respectively. We combine binary members when their separations become smaller than the stellar radius of more massive star. The separation decreases suddenly from $\sim1000~$ to $\sim 100~$AU due to the three body interaction with another protostar (i.e. at 35 kyr in the filamentary cloud). The stellar radius grows gradually with time and the binaries finally merge. Other binary mergers take place in a similar manner.

The survival rate of the protostar in the filamentary cloud is larger than in the spherical cloud. This is due to the filament collapse, in which the typical separations between protostars are larger than those in the disk fragmentation. Furthermore, once a filament fragmentation takes place, the disk mass quickly decreases and the binary separation becomes difficult to shrink ($t\sim 10~$kyr in Fig.~\ref{fig_disc_pics}c). 
Here, we compare our survival rate of protostars with other studies. The formation and evolution of multiple stellar system in the normal Pop~III star formation has been followed by some authors \citep[e.g.][]{Stacy+2016, Greif+2012}. They find that about one third of protostars survive at the end of the calculation. In our calculation, the survival rates at $t=100~\mathrm{kyr}$ are $52$ and $38\%$ in the filamentary and the spherical clouds, respectively. 
The survival rate in the spherical cloud is similar to their results while that in the filamentary cloud is much larger than the results of the previous studies. We stress that their calculated clouds and our spherical cloud show no filamentary collapse. 
This fact implies that the survival rate of the protostars is largely determined by the morphology of the collapsing cloud. 
In this study, the cloud morphology is determined by the tidal field originating from the nearby massive galaxy.

In the filamentary cloud, three ejection events are observed. The ejection is caused by the three body interaction. Meanwhile in the spherical cloud, no protostars are ejected. One reason is that the number of protostars are so small that the chance of three body encounter is much smaller than in the filamentary cloud. The other reason is that the gas rich environment in the spherical cloud suppresses the ejection of protostars (Fig.~\ref{fig_sink_masses}b). We find that when the three body encounter takes place in the spherical cloud, a large amount of gas in the disk is ejected instead of the protostars. In fact, the total gravitational energy released by the binary formation is comparable in both clouds. This energy is transferred to the kinetic energy of the surrounding gas in the spherical cloud.

\section{Radiative feedback effects} \label{sec_radiation_feedback}
We calculate the radiative feedback from the accreting protostars by adopting the simplified stellar evolution model (Section~\ref{sec_SEM}). 
Fig.~\ref{fig_1100002_Xion_maps} shows (a) ionization degree, (b) temperature, and (c) density distributions in the filamentary cloud at $t = 38~$kyr.
Here, the ionization degree is defined as $y_\text{HII}/(y_\text{HI} + y_\text{HII})$, where $y_i$ represents the abundance of the chemical species $i$.
Three protostars are emitting ionizing photons at this snapshot and we label them as protostars A, B, and C. Protostars A and B are ejected from the central disk by three-body interaction and they create compact H~II regions around them whose size is $\sim 0.1~\mathrm{pc}$.  Protostar C remains around the center of the disk. We can see ionization degree reaches $0.01$--$0.1$ around protostar C. The temperature around UV emitting protostars only rises
by a factor of $\lesssim 2$ than the surrounding neutral gas, since the surrounding gas has already a high temperature
(Fig.~\ref{fig_1100002_Xion_maps}b). Thus, the ionizing radiation causes only minor effects on the protostar evolution in our calculations. 

In this section, we investigate the effect of the radiation feedback in detail and discuss whether the radiation feedback
prevents the further mass accretion or not.

\begin{figure}
	\centering
		\includegraphics[width=7.cm]{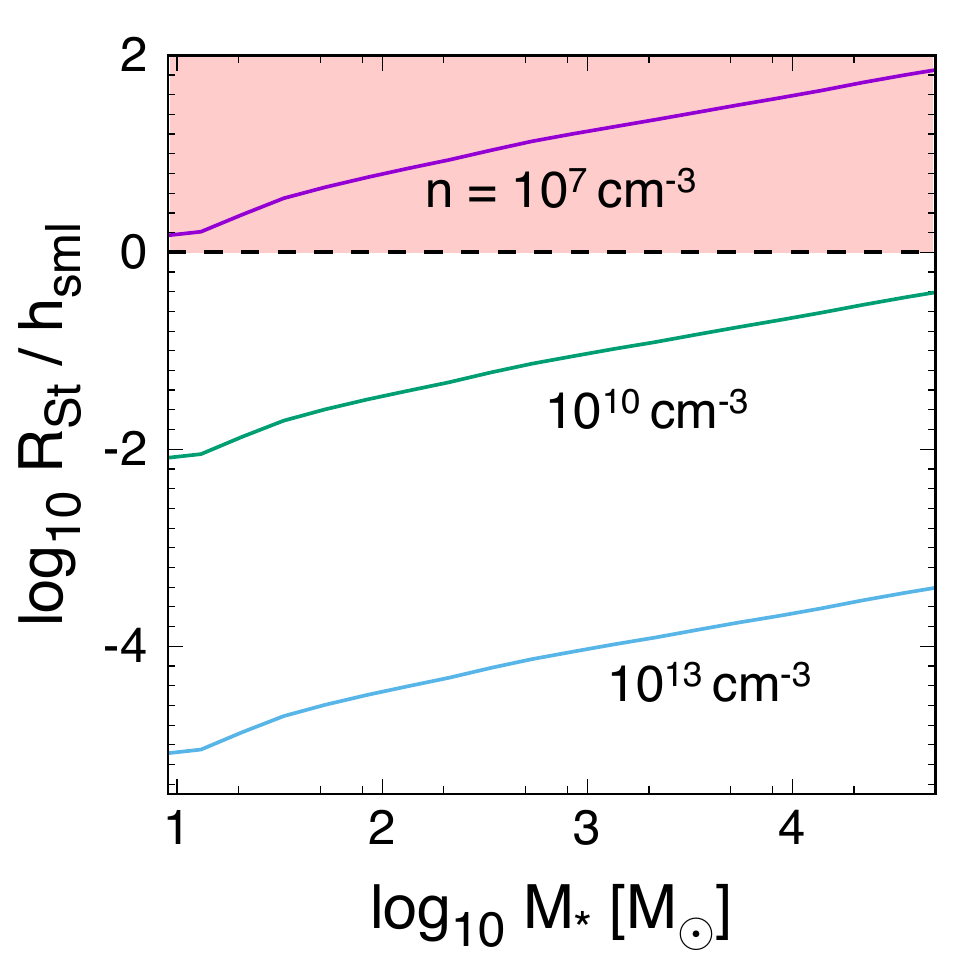}
		\caption{
The ratio of the Str\"omgren radius $R_\text{St}$ to the smoothing length of the gas particles $h_\text{sml}$ as functions of the stellar mass. The different lines represent the different densities of $n=10^7$ (purple), $10^{10}$ (green), and $10^{13}~\mathrm{cm}^{-3}$ (cyan). The horizontal dashed line corresponds to a boundary of $R_\text{St} = h_\text{sml}$, above which the Str\"omgren radius is spatially resolved with more than one smoothing length. When calculating $h_\text{sml}$ with eq.~\eqref{eq_hsml}, different particle masses are used depending on the density, $1.6~M_\odot$ for $n=10^7~\mathrm{cm}^{-3}$, and $9.6\times10^{-3}~M_\odot$ for $n=10^{10}$ and $10^{13}~\mathrm{cm}^{-3}$, the same values as used in the simulations (Section \ref{sec_splitting}).
		} 
		\label{Rst_Hsml}
\end{figure}

\begin{figure*}
	\centering
		\includegraphics[width=14.5cm]{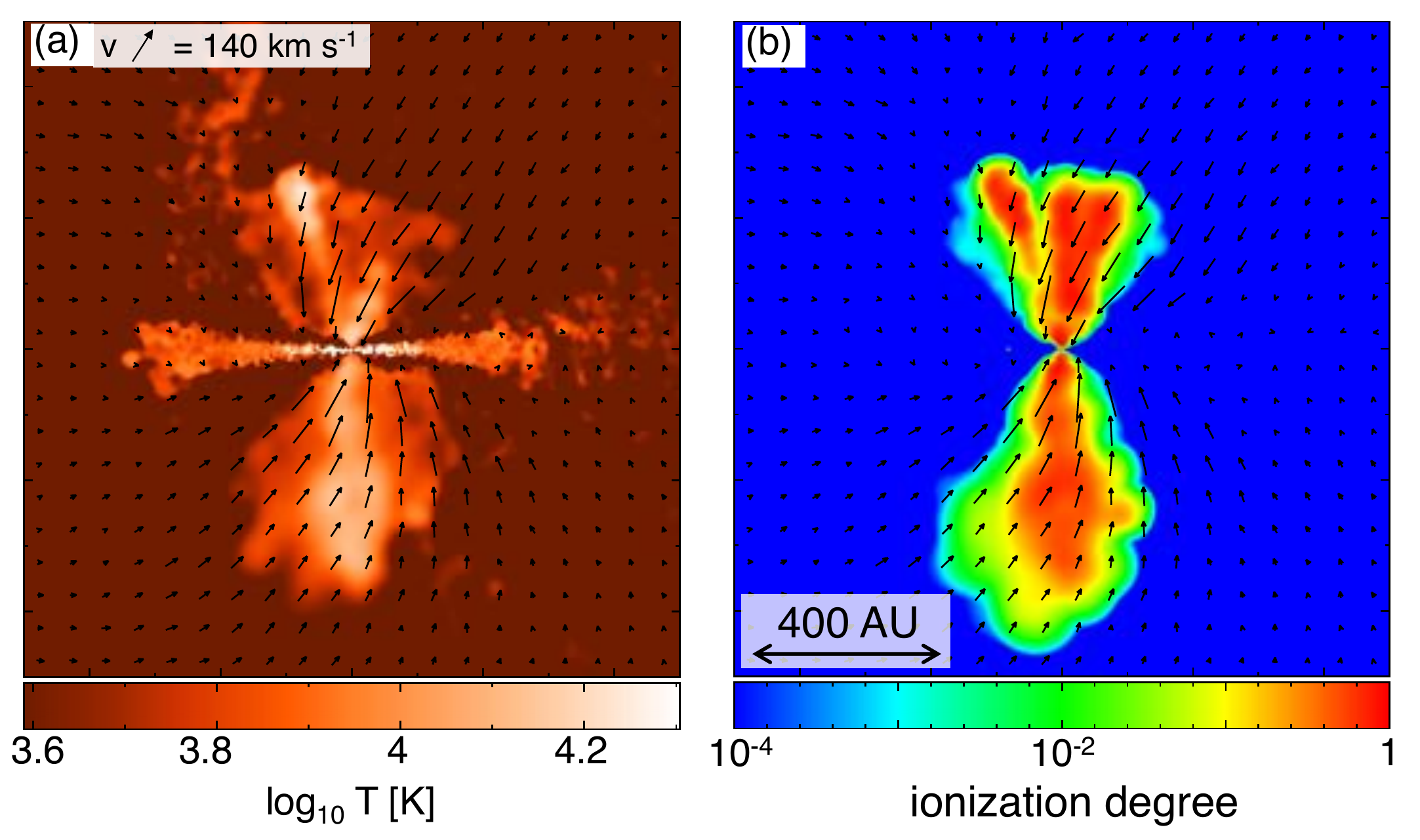}
		\caption{Spatial distributions of the temperature (left) and hydrogen ionization degree (right) around the protostar at 100~years after the particle splitting is performed (see the text). The panels show the edge-on views respect to the circumstellar disk. In each panel, the arrows represent the velocities of the infalling gas. 	} 
		\label{fig_rad_Xion_T_split}
\end{figure*}

\subsection{Limitation of the radiative transfer model}
One can argue that the radiation feedback is inefficient 
because, in a high density region, we cannot resolve the initial Str\"omgren radius ($R_\text{St}$),
within which ionization and recombination balance. The expansion of the ionized region is
mainly driven by the thermal pressure. In order to follow the expansion of the ionized region,
we need to first resolve the structure within $R_\text{St}$. 

If the density ($\bar{n}$) is uniform around the ionizing source, $R_\text{St}$ can be written as,
\begin{align}
R_\text{St} &= \left ( \frac{3 L_\text{UV}}{4\pi\bar{n}^2\alpha_\text{B} E_\text{UV}} \right )^{1/3} \nonumber \\
&= 300~\mathrm{AU} \left ( \frac{S_\text{UV}}{10^{53}~\mathrm{s^{-1}}} \right )^{1/3}
\left ( \frac{\bar{n}}{10^9~\mathrm{cm}^{-3}} \right )^{-2/3},
\end{align}
where $\alpha_\text{B}$ is the case-B recombination coefficient, $L_\text{UV}$ is the ionizing luminosity, $E_\text{UV}$ is the mean energy of the ionizing photons, and $S_\text{UV}$ is the number of the emitted ionizing photons per unit time.
To resolve the initial $R_\text{St}$, the smoothing length of the SPH particles ($h_\text{sml}$) should be sufficiently
smaller than $R_\text{St}$. The smoothing length $h_\text{sml}$ are determined in the simulation as
\begin{align} \label{eq_hsml}
N_\text{neib} m_\text{sph} &= \frac{4\pi h_\text{sml}^3}{3} \mu m_\text{p} n,
\end{align}
where $N_\text{neib} = 64$ is the number of the neighbor SPH particles to determine $h_\text{sml}$, $m_\text{sph}$ is the particle mass,
$\mu$ is the mean molecular weight, and $n$ is the density of the gas particle. Equation \eqref{eq_hsml} shows that $h_\text{sml}$ decreases
as we reduce $m_\text{sph}$. However, since our goal is to follow the long-term evolution, we cannot reduce the particle mass further.

Fig.~\ref{Rst_Hsml} shows the ratios of $R_\text{St}$ to $h_\text{sml}$ for different stellar masses and ambient densities. Here, $S_\text{UV}$ is given by the result of \cite{Hosokawa+2013} considering the star is contracted to ZAMS. In the dense region $n\gtrsim 10^{10}~\mathrm{cm}^{-3}$, $R_\text{St}$ is smaller than $h_\text{sml}$ for the mass range considered. Only in the region with $n \lesssim 10^{7}$--$10^{8}~\mathrm{cm}^{-3}$, $R_\text{St}$ is marginally resolved. 
In Fig.~\ref{fig_1100002_Xion_maps}(c), the ejected protostars A and B lie at the density smaller than $10^8~\mathrm{cm}^{-3}$. This is why the ionized region expands around those protostars. However, protostar C is located inside the disk where $n \gtrsim 10^9~\mathrm{cm}^{-3}$. With the stellar mass $\sim 10^3~M_\odot$, $R_\text{St}$ is smaller than $h_\text{sml}$ of the surrounding gas for this case (Fig.~\ref{Rst_Hsml}). Thus, the ionized region is difficult to expand around this protostar.

\subsection{Radiation transfer with higher resolution}
To further investigate the ionizing radiation feedback, 
we follow the expansion of the ionized regions accurately by adopting a higher spatial resolution.
We focus on the radiation from the most massive protostar in the filamentary cloud.
We increase the spatial (and mass) resolution in the polar region, by splitting
each gas particle into $10^3$ particles. By this procedure, the polar region is filled with a sufficiently large number of gas particles,
and it is possible to resolve $R_\text{St}$ by more than ten times the local smoothing length $h_\text{sml}$.
We follow the expansion of H~II regions for a few hundred years.

Fig.~\ref{fig_rad_Xion_T_split} shows the temperature (left) and $X_\text{ion}$ (right) distributions around the protostar. The arrows indicate
the velocity field of the surrounding gas. The ionized region expands in the polar directions.
The gas in the polar region is heated up to $2 \times 10^4~$K, but the ionized gas continues falling onto the disk, and the expansion of the H~II region is soon halted.
This is because the temperature of this ionized gas is only by a factor of two to three larger than the neutral gas,
and thus is still gravitationally bound by the central star.
We evaluate the gravitational radius ($R_\text{B}$) at which the gas is bound by the central star as 
\begin{align} 
R_\text{B} &= \frac{GM_*}{c_\text{s,HII}^2} \nonumber \\
& = 3.25 \times 10^3~\mathrm{AU} \left ( \frac{M}{10^3~M_\odot} \right ) \left ( \frac{T}{2\times 10^4~\mathrm{K}} \right )^{-1} \left ( \frac{\mu}{0.6} \right ). \nonumber
\end{align}
The protostar mass at this time is about $6000~M_\odot$, and $R_\text{B}$ is much larger than the size of the ionized region, $\sim 400~$AU.
Thus essentially all the ionized gas is gravitationally bound.

We calculate the mass accretion with and without the ionizing radiation for a few hundred years after the particle splitting.
The accretion rates onto the central star show almost no difference between these two runs.
The gas within the disk is strongly bound by the central star and not photo-evaporated.
We confirm that the infall velocity profiles within the polar region with and without the ionizing radiation
remain nearly unchanged and the heating by ionization has little impact on the infalling gas.

Our analytic estimate also shows that the ionized gas will be gravitationally bound until the mass of the central
star exceeds $\sim 10^{5}~M_{\odot}$. Specifically, we construct the density profile toward the polar region and
discuss whether the edge of the ionized region is bound by the central star. More detailed description is
given in the Appendix~\ref{sec_appendix_nprofile}. Our analysis there supports the simulation result;
the ionizing radiation has negligible impact on the accreting matter.

\section{Discussion} \label{sec_discussion}
\subsection{Final Mass and Fate of the SMS}
Although we have followed the protostellar evolution for $\sim 0.1~$Myr, it is still way before an SMS collapses into a BH. In order to determine how massive seed BHs are finally left, we need to follow the further evolution for the stellar lifetime $\sim$ 2 Myr.
However, it is computationally too expensive to accomplish it. 
We here estimate the final stellar and BH masses from the final snapshots in our simulations. 

\begin{figure}
	\centering
		\includegraphics[width=8.cm]{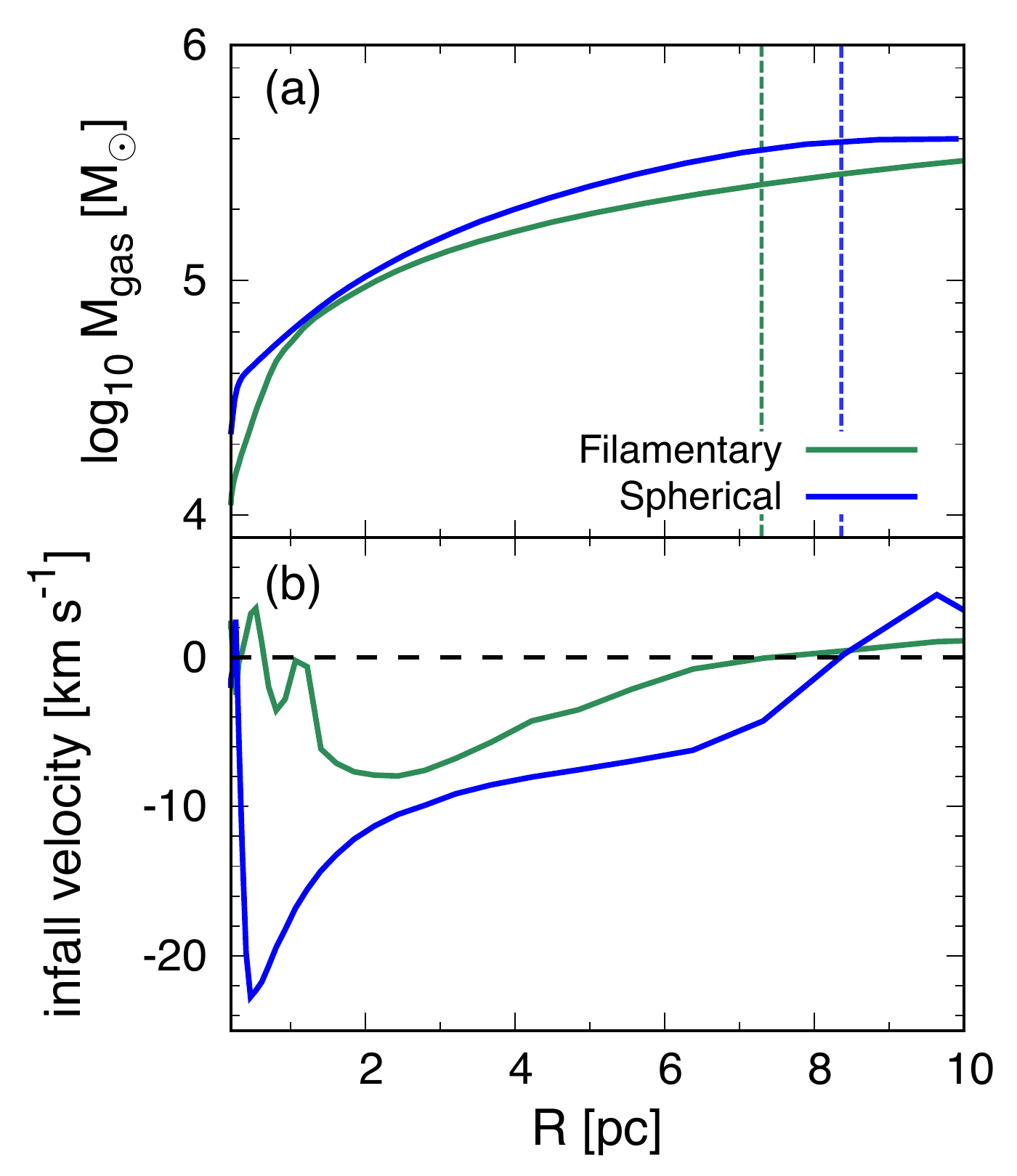}
		\caption{Radial profiles of (a) enclosed gas mass and (b) infall velocity at $0.1~$Myr after the birth of the primary protostar for the filamentary (green) and spherical (blue) clouds. The horizontal axis represents the distance ($R$) from the most massive protostar. In panel (a), the green and blue vertical dashed lines mark the outer boundaries of the clouds, beyond which the gas is moving outward. In panel (b), the horizontal dashed line shows the boundary between the inflow and outflow. Note that a positive value represents the outflow in this plot. 		}
		\label{fig_infall_vel}
\end{figure}

Fig.~\ref{fig_infall_vel} shows the radial profiles of (a) the enclosed gas mass and (b) the gas velocity at the final epoch of our simulations
for the filamentary (green) and the spherical (blue) clouds. We see that the gas is outflowing at $R > 7$--$8~$pc in both clouds, which is caused by
tidal disruption by the nearby massive galaxy. The infalling gas within $R < 7$--$8~$pc can reach the central star
within the stellar lifetime, $\sim$ 2 Myr. Panel (a) shows the infalling gas mass is $\lesssim 3\times 10^5~M_\odot$ for both clouds.

Assuming that no more protostars appear via the fragmentation, and that the existing protostars equally accrete the infalling gas, 
we can estimate the typical final masses to be $\sim 10^4~M_\odot$ for the filamentary cloud and $\sim 10^5~M_\odot$ for the spherical cloud.
These stars are massive enough to collapse into BHs after exhaustion of nuclear fuel \citep[e.g.][]{Umeda+2016,Woods+2017}.
During the final gravitational collapse, most of the stellar mass is swallowed by the BH \citep[e.g.][]{Shibata+2016,Uchida+2017}.
Therefore, we expect that the filament cloud yields about ten BHs with $\sim 10^4~M_\odot$ and
the spherical cloud produces several BHs with $\sim 10^5~M_\odot$. In what follows we call these BHs as ``DCBHs''.

\subsection{Evolution of the DCBH binaries} \label{sec_discussion_binary_evo}
We find that the stellar binaries are formed in the cloud core. Some of them survive until the end of the simulation, without merging with the companion.
After their stellar lifetime, they will evolve into massive BH binaries
with masses of $10^3$--$10^5~M_\odot$ and with separations of $10^2 \sim 10^3~$AU (see Fig.~\ref{fig_binary_props}). 
One process for the binaries to lose their angular momenta and to coalesce
is emission of gravitational waves. The coalescence time is given by \citep{Peters1964};
\begin{equation}
t_\text{GW,merge} = 1.25\times 10^{11} ~\mathrm{yr} \left (\frac{a}{100~\mathrm{AU}} \right )^4 \left ( \frac{M_\text{BH}}{10^5~M_\odot} \right )^{-3},
\end{equation}
which is about an order of magnitude larger than the Hubble time. 
Some additional processes to remove the angular momentum, e.g., interactions with surrounding gas or stellar components, are thus necessary for a BH-BH merger to occur.

If the gas or stars are accreted onto a BH, a fraction of the accreted material is scattered and
carries away the angular momentum from the system. \cite{Kashiyama&Inayoshi2016} estimated the number of stars
falling onto the central BH, assuming possible star cluster formation. 
In our simulations, clusters with several to ten stars are formed.
Following \cite{Kashiyama&Inayoshi2016}, the relaxation time of these cluster ($t_\text{relax}$) is,
\begin{align}
t_\text{relax} & \sim 1.6 \times 10^5~\mathrm{yr} \frac{M_\text{BH}}{10^5~M_\odot} \frac{10^3~M_\odot}{\langle M_* \rangle} \left ( \frac{r}{\mathrm{pc}} \right )^{-3/2},
\end{align} 
where $M_\text{BH}$ is the central BH mass,  $\langle M_* \rangle$ is the average mass of formed stars,
and $r$ is the size of the star cluster. Cluster member stars are scattered into the loss cone orbit
with the time scale of $t_\text{relax}$ and falls onto the central BH.
If the central object is a BH binary instead, then a part of the accreted star will be ejected.
If we assume all the accreted star ejected at a speed of the escape velocity $v_\text{esc} = \sqrt{GM_\text{BH}/a}$,
then the angular momentum loss with time $t_\text{star,merge}$ is
\begin{align}
&t_\text{star,merge} \sim \frac{J}{\langle M_* \rangle v_\text{esc} a / t_\text{relax}} \sim \frac{M_\text{BH}}{\langle M_* \rangle } t_\text{relax}  \nonumber \\
&\;\;\; \sim 1.6 \times 10^7~\mathrm{yr} \left ( \frac{10^3~M_\odot}{\langle M_* \rangle} \right )^2 \left ( \frac{M_\text{BH}}{10^5~M_\odot} \right )^2 \left ( \frac{100~\text{AU}}{a} \right ).
\end{align}
Thus, the BH binary separation can decrease through interaction with the stars formed in the same cloud.

After the formation of the DCBH, the host cloud will merge with the nearby massive galaxy (the LW radiation source).
The stellar mass of the source galaxy is $10^6$ and $10^7~M_\odot$
for the filamentary and the spherical clouds, respectively. The mass of the cold gas within the source galaxy
is an order of magnitude larger than the stellar mass.
The BH binary is likely to shrink by interacting with the cold gas and stars inside the source galaxy,
and will finally merge due to gravitational wave emission.

\begin{figure}
	\centering
		\includegraphics[width=7.5cm]{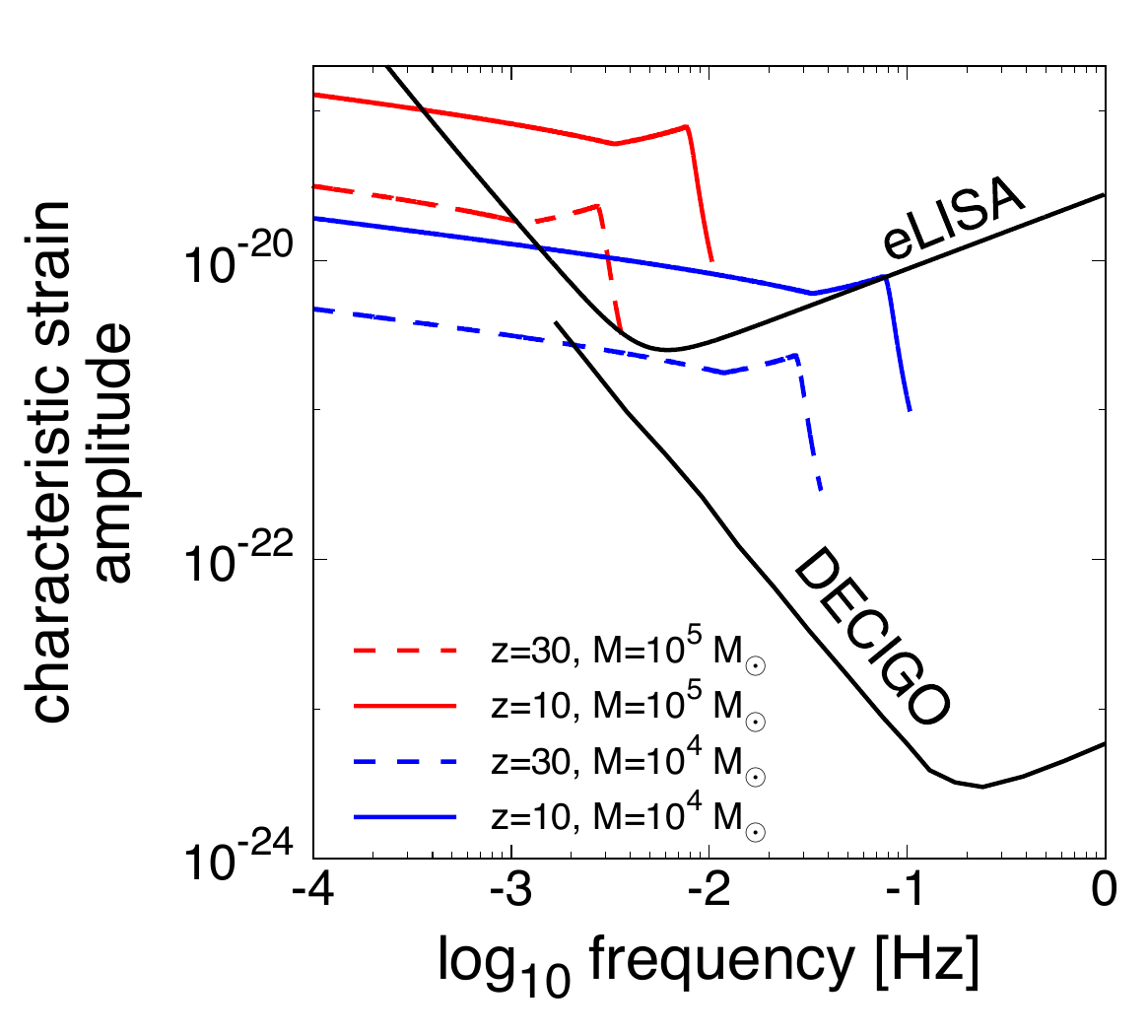}
		\caption{Characteristic strain amplitudes for merging BH-BH binaries against noise amplitudes of the future observational facilities. Only the equal-mass binaries with $10^4~M_\odot$ BHs (blue) and $10^5~M_\odot$ BHs (red) are considered. For each case, the solid and dashed lines represent the merger events occurring at $z=10$ and $30$. The black solid lines show the expected noise amplitudes of eLISA \citep{Amaro-Seoane+2012} and DECIGO \citep{Kawamura+2011}
		} 
		\label{fig_GW}
\end{figure}

The amplitudes of the gravitational waves from merging DCBHs peak at the frequency of $1$--$10~$mHz at the rest frame.
Thus the ground based GW detectors are difficult to detect the GWs.
The space GW detectors such as evolved Laser Interferometer Space Antenna (eLISA) and Deci-hertz Interferometer Gravitational 
wave Observatory (DECIGO) can target these GWs. 

Fig.~\ref{fig_GW} shows the characteristic strain amplitude $h_c(f)$ for binaries at $z=10$ (solid) and $30$ (dashed).
The waveforms are calculated following the template given by \cite{Ajith+2011}.
We assume equal mass binaries, where the mass of each BH is $10^4$ (blue) and $10^5~M_\odot$ (red).
Black solid lines show the noise amplitudes of eLISA and DECIGO. 
Binaries with $10^4$--$10^5~M_\odot$ can be observed at $z \gtrsim 10$.
If the seed BH of the observed SMBHs are provided by the DC model, we expect a large number of BH binaries
with $10^4$--$10^5~M_\odot$ at $z > 6$.
Although the exact number density of DCBHs is under debate,
detection of gravitational waves from DCBH binaries in this mass range will place important
constraints on the formation scenario of SMBHs.

\begin{figure}
	\centering
		\includegraphics[width=8.cm]{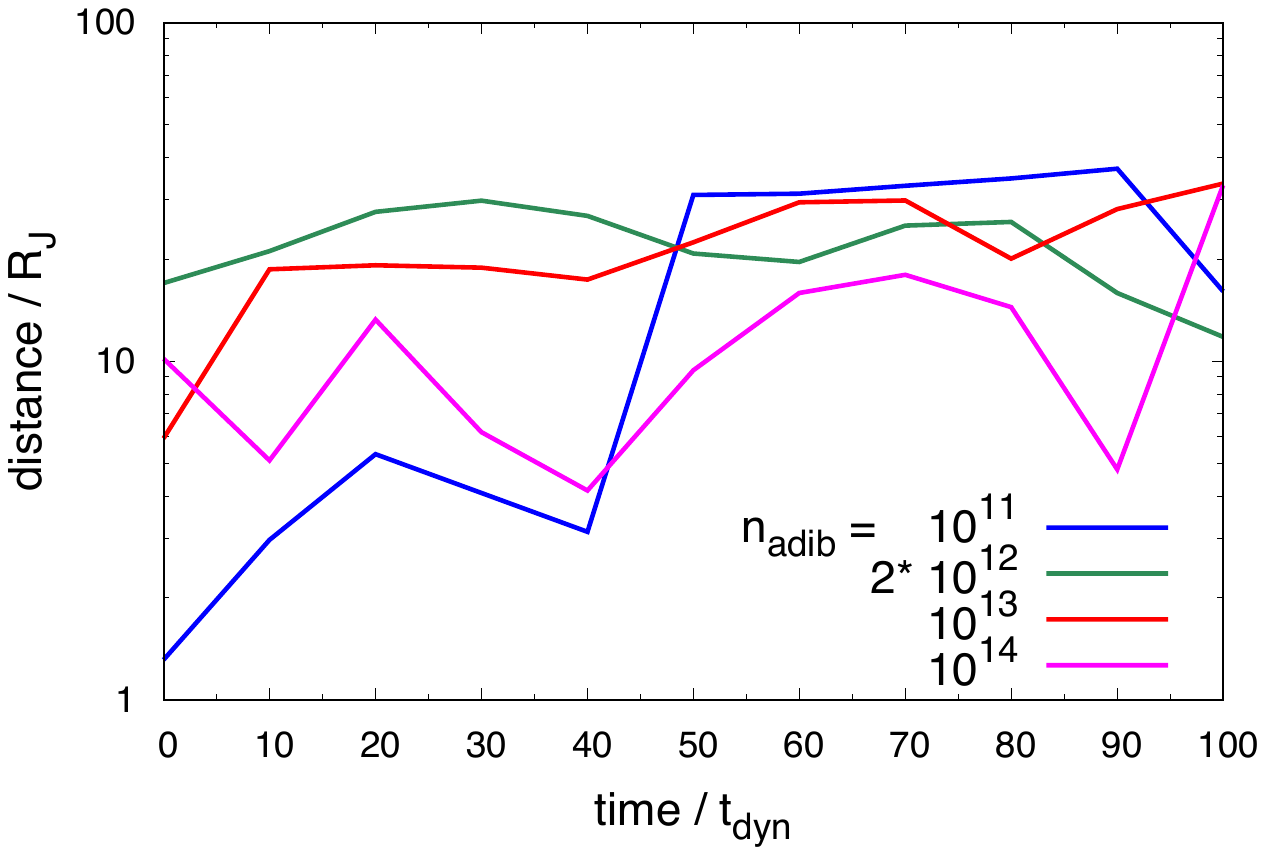}
		\caption{Time evolutions of the separation between the most massive protostar and the closest one with different threshold number densities $n_\text{adib}$, above which the gas is assumed to evolve adiabatically. The time is measured in the unit of the free-fall time with the threshold density $n_\text{adib}$ for each case. The separation is normalized by the Jeans length with the same threshold density and the fixed temperature $T=8000$~K. The time origin corresponds to the moment when the closest protostar is formed in the disk.} 
		\label{fig_Resolution}
\end{figure}

\subsection{Effect of FUV self-shielding on the disk structure} \label{sec:FUV_rad}
Since we do not include H$_2$ self-shielding against the external LW radiation 
in this study, we may overestimate the photodissociation rate. 
To see whether the shielding effect can modify the disk structure
with some enhanced H$_2$ abundance, we have performed following test simulations; for simplicity, we completely omit the LW field when the gas density exceeds $n_\text{crit}=10^6~\mathrm{cm}^{-3}$, where the cloud becomes optically thick to the external LW radiation \citep{Draine+1996}. 
We have continued to follow the evolution in the accretion stage for 2000 years. 

As a result, we find almost no differences between the simulations 
with and without the external LW field.
Even without the LW field, H$_2$ is completely destroyed within the disk, typically only with the H$_2$ abundance of $\sim10^{-8}$. The atomic hydrogen cooling still dominates the molecular cooling. 
Some regions are cooled by adiabatic expansion to have $\sim 3000~$K,
but it is also seen in the default cases with the LW field.
We thus conclude that the presence of the LW field has almost no impacts on the disk structure. Our ignorance of the FUV self-shielding has no effects either. 

In fact, H$_2$ molecules are mainly destroyed by collisional dissociation, not by photodissociation with LW photons. \cite{Inayoshi+2012} show that it occurs in the thermal state characterized by $n \gtrsim 10^4~\mathrm{cm^{-3}}$ and $T \gtrsim 5000$~K. In our simulations, most part of the disk really satisfies the condition for the collision to dominate the dissociation.

\subsection{Numerical resolution} 
\label{sec:resolution}

In our study, the spatial resolution is limited by the adiabatic threshold density $n_\text{adib}$, above which the gas is forced to behave adiabatically with $\gamma = 5/3$. 
In reality, the threshold density is physically determined as $n_\text{adib} \simeq 10^{16}~\mathrm{cm}^{-3}$ \citep{Omukai2001}.  
The reason why we use the lower value $n_\text{adib}=10^{13}~\mathrm{cm}^{-3}$ is to save computational costs to follow the long-term evolution. To see how such an artificial EOS affects our results, we run test simulations with changing the threshold density.

The test simulations follow the evolution in the accretion stage where the disk fragmentation occurs in the filamentary cloud.
In particular, we focus on the evolution of the orbital separation $R_\text{sep}$
between the most massive protostar and another one close to it. 
Fig.~\ref{fig_Resolution} shows the time evolution of $R_\text{sep}$ with different thresholds $n_\text{adib} = 10^{11}$ (blue), $2\times 10^{12}$ (green), $10^{13}$ (red), and $10^{14}~\mathrm{cm}^{-3}$ (magenta). We here normalize the time by the dynamical time $t_\text{dyn}$ and the separation by the Jeans length $R_\text{J}$ assuming a fixed temperature of $8000~$K, 
\begin{align}
t_\text{dyn} &= 27~\mathrm{yr} \left( \frac{n_\text{adib}}{10^{13}~\mathrm{cm}^{-3}} \right )^{-1/2},\\
R_\text{J}   &= 44~\mathrm{AU} \left( \frac{n_\text{adib}}{10^{13}~\mathrm{cm}^{-3}} \right )^{-1/2} \left( \frac{T}{8000~\mathrm{K}} \right )^{1/2}.
\end{align}

In all the examined cases, the disk violently fragments into many pieces in an early stage of $t \lesssim 50~t_\text{dyn}$. Some protostars migrate inward to merge with the central star, whereas others remain in stable orbits. Fig.~\ref{fig_Resolution} shows that the separation converges to a few ten times $R_\text{J}$ after $\simeq 50~t_\text{dyn}$ in all the cases. This suggests that more tighter binaries will be resolved with the higher threshold density, with which the Jeans length is smaller and resulting resolution is higher \citep{Machida&Doi2013}.

In our cases, however, recall that the radius of a rapidly accreting protostar with $\sim 10^3~M_\odot$ is $\simeq 40~$AU (see eq.~\ref{eq_rstar}) and tight binaries with the smaller separations are assumed to be merged. 
The Jeans length with the default threshold density $n_\text{adib}=10^{13}~\mathrm{cm}^{-3}$ is $R_\text{J} \simeq 40~$AU, comparable to the radius of the bloated protostars. Binaries with the smaller separations will appear with the higher threshold, but in any rate they will be merged away after the stars grow in mass and greatly inflate. 
We thus conclude that our choice of somewhat low threshold density will only have a limited effect on our conclusions. It is to be confirmed by future simulations with substantially higher spatial resolutions.

\section{Summary} 
\label{sec_conclusion}

We have studied long-term ($\sim 0.1~$Myr) evolution through the accretion 
phase of SMS formation by performing 3D radiation-hydrodynamic simulations.  
We argue that tidal force of nearby galaxies plays a critical role.
As shown by \cite{Chon+2016}, first of all, the cloud collapse is actually prevented by this effect in many cases. In this paper, we show that the tidal force still dominates the later evolution for the rare cases where the cloud collapses.  
In our simulation, one cloud is distorted by the strong tidal force of a nearby massive galaxy,
and is deformed to be a filament. Fragmentation yields multiple star-disk systems,
and further fragmentation occurs in each disk, and more than 10 stars
with a few $\times~10^3~M_\odot$ are finally formed.
In the other case, the gas cloud collapses almost spherically with a relatively weak tidal field. 
Only a single star-disk system forms in this case, and a few stars are formed
within the disk. The stellar masses at $0.1$~Myr are $\sim \text{a few} \times 10^4~M_\odot$
in this case, but the stars are still accreting at a high rate of $0.1~M_\odot~\mathrm{yr}^{-1}$. 
They are likely to evolve into the SMSs with $\sim 10^5~M_\odot$ before the end of their lives.


Throughout our simulations, the UV feedback plays a minor role. 
There are two reasons for this. 
Firstly, the accretion rate onto a protostar is still as large as $\sim 0.1~M_\odot~\mathrm{yr}^{-1}$,
even in the case where the surrounding gas is accreted onto multiple protostars.
The protostars are in the supergiant stage with having
a cool atmosphere ($T_{\rm eff} \simeq 5000$~K). The resulting UV emissivity is not strong enough
to ionize the surrounding gas.
Secondly, even if the gas is ionized, the effect of the UV feedback is actually limited.
The H~II region remains small and is trapped within the accretion flow.
Since such an H~II region cannot dynamically expand, the accretion is not halted.
According to our analytic estimates, 
an H~II region begins to dynamically break through the accretion envelope 
only after the stellar mass exceeds $\sim 10^5~M_\odot$.


The accreting SMSs in our simulations are expected to collapse into massive BHs
at the end of their lives \citep{Umeda+2016}.
We thus suggest that $10^{4}$--$10^{5}~M_\odot$ BHs are
formed sufficiently early at $z = 18-12$.
Since these BHs will merge into the nearby massive galaxy
and sink toward the galaxy center, they will experience large amount of mass accretion.
If efficient accretion with nearly the Eddington rate lasts until $z\sim7$,
they can attain masses greater than $3 \times 10^8~M_\odot$ by $z = 7$.
Whether radiation from the BH accretion disk prevents the efficient mass accretion
is unclear, and hence warrants further study.

We also find that binaries of $10^4$--$10^5~M_\odot$ SMSs (and hence BHs) are formed 
in both the filamentary and the spherical clouds. 
These binaries will coalesce and emit a huge amount of GWs, which can be detected
by the future space observatories, eLISA and DECIGO.
Such future GW detection will 
provide us a valuable opportunity to understand the formation of early SMBHs.
\\ \\

We thank G. Chiaki, H. Susa, K. Omukai and S. Hirano for fruitful discussions and comments. This work is financially supported by Advanced Leading Graduate Course for Photon Science program, by Grant-in-Aid for JSPS Fellows (16J07507: S.C.), and by the Grants-in-Aid for Basic Research by the Ministry of Education, Science and Culture of Japan (15H00776, 16H05996, 17H06360: T.H.; 25287050: N.Y.). Numerical computations are carried out on XC30 at the Center for Computational Astrophysics (CfCA) of the National Astronomical Observatory of Japan. We use the SPH visualization tool SPLASH \citep{SPLASH} in Figs. \ref{fig_LSS}, \ref{fig_density_maps_1100002}, \ref{fig_disc_pics}, \ref{fig_filament_fragmentation}, \ref{fig_1100002_Xion_maps}, and \ref{fig_rad_Xion_T_split}.

\bibliography{biblio2}

\newpage
\appendix
\newpage

\section{Analytic estimate of the radiation feedback effects} \label{sec_appendix_nprofile}

We here analytically investigate whether the ionizing radiation from the protostar evacuates the surrounding gas. We consider a central star associated with a gas disk. 
As shown in our simulations, a photoionized region first expands toward polar regions where the density is relatively low (Fig.~\ref{fig_rad_Xion_T_split}). 
We first model the density profile along the polar axis by the following density profile,
\begin{align} \label{eq_profile}
n(r) =\left\{ \begin{array}{ll}
\frac{c_\text{s,I}^2}{G\mu_\text{I} m_\text{p}} r^{-2} & (r > R_\text{B,I}) ,\\
\frac{c_\text{s,I}^2}{G\mu_\text{I} m_\text{p}} R_\text{B,I}^{-2} \left ( \frac{r}{R_\text{B,I}} \right )^{-\alpha} & (r < R_\text{B,I}),\\
\end{array}
\right .
\end{align}
where $R_\text{B} \equiv GM_* / c_\text{s}^2$ is the gravitational radius, $c_\text{s}$ the sound speed, and $1 < \alpha < 1.5$ a free-parameter to characterize the inner density profile. 
The physical quantities with subscripts I and II correspond to those
in neutral and ionized regions. We have assumed that the density
profile follows $n \propto r^{-2}$ in the outer region $r > R_\text{B,I}$ 
and it becomes shallower within $R_\text{B,I}$ 
because of the gravity of the central star \citep{McKeeTan2008}.
The spherically collapsing cloud yields $\alpha = 1.5$,
while $\alpha$ slightly decreases when the disk forms around the central star.
In fact, our simulations suggest $1.1 \lesssim \alpha \lesssim 1.3$. 


The Str\"omgren radius $R_\text{St}$, inside which the ionizing photon supply
is fully consumed by the recombination, is calculated by solving
\begin{equation}
S_\text{UV} = \int^{R_\text{St}}_{R_*} 4\pi r^2 \alpha_\text{B} n(r)^2 \mathrm{d}r ,
\end{equation}
where $S_\text{UV}$ and $R_*$ are the stellar UV emissivity and radius. 
With the density profiles given by eq. \eqref{eq_profile}, we obtain
\begin{align} \label{eq_Rst_profile}
R_\text{St} = \left\{ \begin{array}{ll}
&\left [ \frac{(3-2\alpha) G^3\mu_\text{I}^2 m_\text{p}^2 M_*}{4\pi \alpha_\text{B} c_\text{s,I}^6} S_\text{UV} \right . \\
& \;\;\;\;\;\;\;\;\;\;\;\;\;\; \left . + \left ( \frac{R_*}{R_\text{B,I}} \right ) ^{3-2\alpha} \right ]^{1/(3-2\alpha)}
R_\text{B,I} \\
&\qquad \qquad \qquad \qquad \qquad \qquad (\alpha < 1.5), \\
&\\
&R_* \exp \left( \frac{G^3\mu_\text{I}^2 m_\text{p}^2 M_*}{4\pi \alpha_\text{B} c_\text{s,I}^6} S_\text{UV} \right ) \\
&\qquad \qquad \qquad \qquad \qquad \qquad (\alpha = 1.5).
\end{array}
\right .
\end{align}
Note that the expression with $\alpha = 1.5$ coincides with that found in \cite{OmukaiInutsuka2002}.
We assume $S_\text{UV}$ is equal to the Eddington luminosity, which is valid for very massive stars, and $R_*$ as the ZAMS radius,
\begin{align}
S_\text{UV} &= 10^{49} ~\mathrm{s^{-1}} \left ( \frac{M*}{M_\odot} \right ), \\
R_* &= 4.24 ~R_\odot \left ( \frac{M_*}{100~M_\odot} \right )^{0.59}.
\end{align}
To see whether the ionized gas is gravitationally bound, we compare the Str\"omgren radius to
the gravitational radius of the ionized gas. For $\alpha < 1.5$, we get
\begin{align} \label{eq_Rbondi_profile}
\frac{R_\text{St}}{R_\text{B,II}} &= \left (\frac{c_\text{s,II}}{c_\text{s,I}} \right )^2 \left [ \frac{(3-2\alpha) G^3 \mu_\text{I}^2 m_\text{p}^2 S_\text{UV} M_*}{4\pi \alpha_\text{B} c_\text{s,I}^6}  \right ]^{1/(3-2\alpha)} \nonumber \\
&= 3 \left [ 4.4\times10^{-5} (3-2\alpha) \left (\frac{M_*}{10^3~M_\odot}\right )^2 \right ]^{1/(3-2\alpha)},
\end{align}
where we have neglected the second term of eq.~\eqref{eq_Rst_profile}. 
In the last equation, we assume $\mu_\text{I} = 1.2$, $\mu_\text{II} = 0.6$, $T_\text{I} = 8000~$K, and $T_\text{II} = 1.5\times 10^4~$K, which are motivated by our simulation results.

\begin{figure}
	\centering
		\includegraphics[width=8.cm]{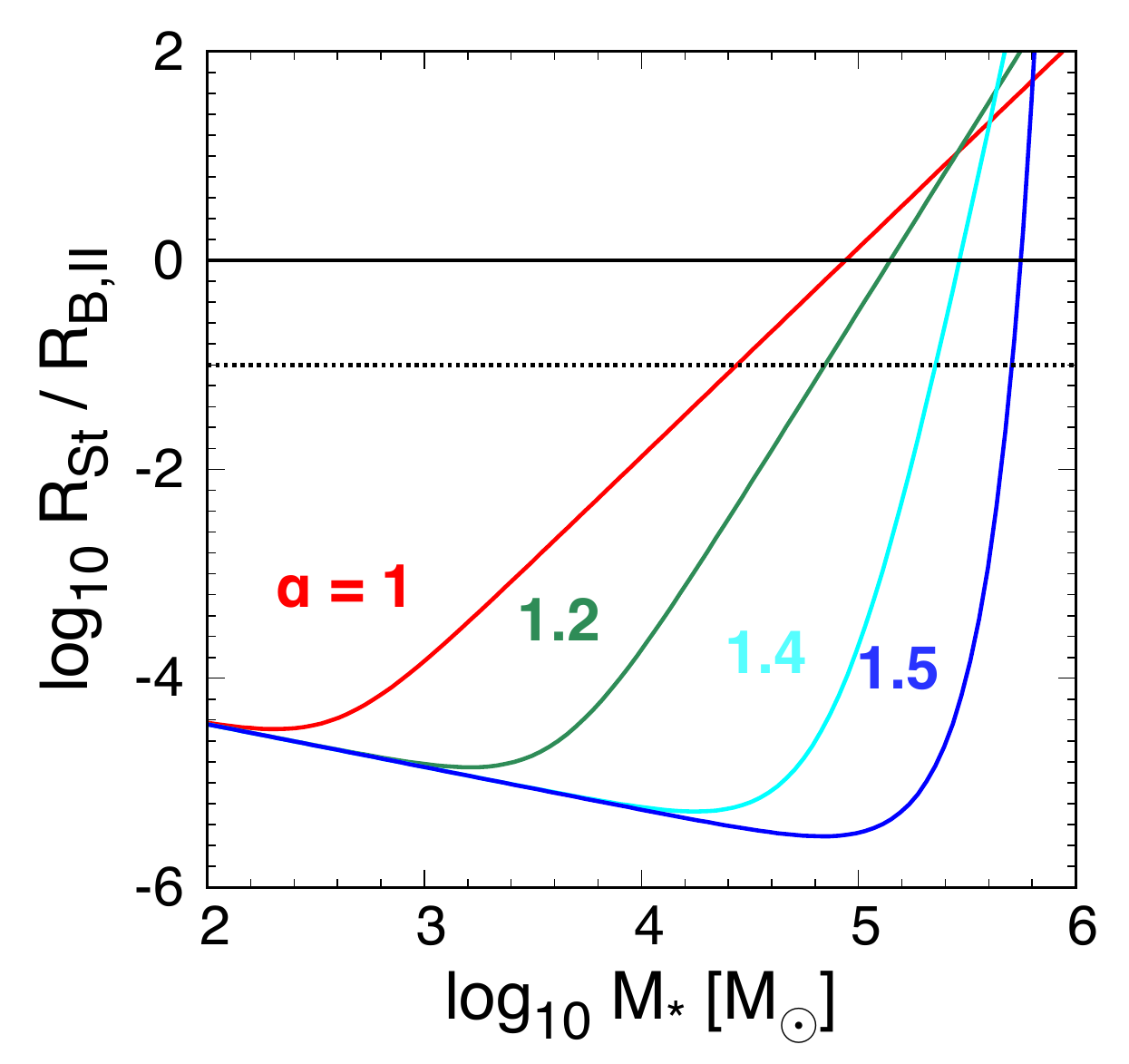}
		\caption{Ratios of the Str\"omgren radius $R_\text{St}$ (eq.~\ref{eq_Rst_profile}) to gravitational radius of the ionized gas $R_\text{B,II}$ as a function of the mass of the central star. The red, green, cyan, and blue lines represent the different density profiles within the gravitational radius (eq.~\ref{eq_profile}) with $\alpha=1$, $1.2$, $1.4$, and $1.5$.
The horizontal solid line is a critical line of $R_{\rm st} = R_{\rm B,II}$, above which a photoionized region extends beyond the gravitational radius. The horizontal dotted line is also another critical line for the breakout of the ionized region $R_{\rm st} = 0.1 R_{\rm B,II}$, considering the effect of the radiation pressure against the gravity (see text).
} 
\label{fig_Rst_to_Rbondi}
\end{figure}

Fig.~\ref{fig_Rst_to_Rbondi} shows $R_\text{St} / R_\text{B,II}$ 
as a function of $M_*$ for different $\alpha$. 
For $1 < \alpha < 1.5$, the Str\"omgren radius exceeds the gravitational radius only for $M_* \gtrsim 10^5~M_\odot$. The ionized region is gravitationally bound and confined around the star until the stellar mass exceeds $\sim 10^5~M_\odot$.
Such a trapped ionized region never disturbs the mass accretion. We thus reinforce our argument in Section~\ref{sec_radiation_feedback}, i.e., the UV feedback plays almost no roles in the evolution followed in our simulations.


We have neglected the effect of the radiation pressure in the above estimate.
Thomson scattering effectively counteracts the gravity, so that it helps the 
expansion of the ionized region. 
Such an effect is included in our modeling with reducing $R_\text{B,II}$ by a factor of $\Gamma \equiv 1 - L/L_\text{Edd}(M_\text{tot})$, where $L$ is the stellar
luminosity, $L_\text{Edd}$ the Eddington luminosity, and $M_\text{tot}$ the total mass within the ionized region including the gas and star. Accordingly, the condition for the breakout of the ionized region is modified as $R_\text{St} / R_\text{B,II} = \Gamma$. 
The highest-mass stars that appear in our simulations have $\Gamma \sim 0.1$, but Fig.~\ref{fig_Rst_to_Rbondi} still shows that the breakout occurs only for $M_* \gtrsim 10^4~M_\odot$.
Therefore, our conclusions are not affected by this effect. 

\end{document}